\documentclass[preprint]{elsarticle}
\usepackage{amsmath}
\usepackage{amssymb}
\usepackage{epsfig}
\usepackage{epstopdf}
\usepackage[latin5]{inputenc}
\usepackage{subfigure}
\usepackage{lineno,hyperref}

\modulolinenumbers[5]

\usepackage{color}

\numberwithin{equation}{section}

\begin{document}

\title{Dispersive shock waves in the Kadomtsev-Petviashvili and Two Dimensional Benjamin-Ono equations }

\author[a]{Mark J. Ablowitz}
\ead{mark.ablowitz@colorado.edu}

\author[a,b]{Ali Demirci\corref{cor1}}
\ead{demircial@itu.edu.tr}

\author[a]{Yi-Ping Ma}
\ead{yiping.m@gmail.com}

\cortext[cor1]{Corresponding author}
\address[a]{Department of Applied Mathematics, University of Colorado, 526 UCB, Boulder, CO 80309-0526, USA}
\address[b]{Department of Mathematics, Faculty of Science and Letters, Istanbul Technical University, Istanbul 34469, Turkey}

\begin{abstract}
Dispersive shock waves (DSWs) in the Kadomtsev-Petviashvili (KP) equation and two dimensional Benjamin-Ono (2DBO) equation are considered using parabolic front initial data. Employing a front tracking type ansatz
exactly reduces the study of DSWs in two space one time $(2+1)$ dimensions to finding DSW solutions of  $(1+1)$ dimensional equations. With this ansatz, the KP and 2DBO equations can be exactly reduced to  cylindrical Korteweg-de Vries (cKdV) and  cylindrical Benjamin-Ono (cBO) equations, respectively.
Whitham modulation equations which describe DSW evolution in the cKdV and cBO equations are derived in general
and Riemann type variables are introduced. DSWs obtained from the numerical solutions of the corresponding Whitham systems and direct numerical simulations of the cKdV and cBO equations are compared with excellent agreement obtained.
In turn, DSWs obtained from direct numerical simulations of  the KP and 2DBO equations are compared with the cKdV and cBO equations, again with remarkable agreement. It is concluded that the $(2+1)$ DSW behavior along parabolic fronts can be effectively described  by the DSW  solutions of the reduced $(1+1)$ dimensional equations.

\end{abstract}


\date{\today}

\maketitle

\section{Introduction}

In recent years the study of dispersive shock waves (DSWs) has generated considerable interest. In water waves DSWs have also been termed undular bores \cite{Ligh78} \cite{Smy88}.  In fact, an early observation of an undular bore goes back to 1850 \cite{Baz50}. In plasma physics a careful observation of a DSW, sometimes referred to as  a collisionless shock wave, was made over 40 years ago \cite{Tay70}. More recent experiments/observations of DSWs have occured in other fields, e.g. Bose-Einstein condensates (BEC) \cite{Hoe06, Hoe08} and nonlinear optics \cite{Wan07, Con09, Fat14}. Mathematically speaking the study of DSWs is difficult since the profile of the shock wave is highly oscillatory and the underlying shock solution does not converge strongly. A prototypical  example of a DSW occurs in the KdV equation
\begin{equation}
\label{kp}
u_t+uu_x+\epsilon^2 u_{xxx}=0
\end{equation}
with  $\epsilon^2 \ll 1$ and inital conditions corresponding to a simple unit step (Heaviside) function.  In 1974, employing an averaging  method pioneered by Whitham \cite{Whi65}, Gurevich and Petiavskii \cite{Gru74} gave a detailed description of the associated DSW. About 10 years later Lax and Levermore \cite{Lax83}
described the DSW rigorously via inverse scattering transform methods. Over the years there have been numerous important analytical  studies that employ Whitham methods cf. \cite{Gru74, Gru87, GA07, Mat98, Mat07}.

Here we study two space one time $(2+1)$ dimensional equations, including the Kadomtsev-Petviashvili (KP) \cite{KP70}
and the two dimensional Benjamin-Ono (2DBO) \cite{Abl80} equations (see Eqs.~(\ref{kp}) and (\ref{kpbo})), by exactly reducing these equations to the $(1+1)$-dimensional cylindrical KdV (cKdV) and cylindrical Benjamin-Ono (cBO) equations (see Eqs.~(\ref{cKdV}) and (\ref{cBO})) respectively.

We  analyze the cKdV/cBO equations via Whitham  theory and derive the general Whitham modulation equations; these equations are transformed into simpler form by introducing appropriate Riemann type variables. These Whitham equations in Riemann variables are not  in diagonal form. We remark that in the cKdV case diagonal form may be obtained using the integrability of cKdV \cite{Abl91, Cal78, John79}; on the other hand, neither 2DBO nor its reduction, the cBO equation is known to be integrable.

We study the DSWs in the cKdV and cBO equations numerically and describe their differences from the DSWs in the classical KdV and BO equations. Indeed the DSWs in the former  are found to decay slowly in time whereas those in the latter do not exhibit such temporal decay. We find that direct numerical simulations of the Whitham modulation equations agree well with those of the cKdV and cBO equations.

We then compare these $(1+1)$ dimensional DSW structures to direct numerical simulations of the $(2+1)$ dimensional KP and 2DBO equations. Our comparisons between $1+1$ numerics/theory and $2+1$ numerics exhibit excellent results. In general the DSW  weakens across the parabolic front as time increases.
We also note that the numerical simulations of $1+1$ Whitham theory which removes the fast variation, are much faster than the $1+1$ cKdV/cBO equations which in turn, are orders of magnitude faster than the $2+1$ equations.

Over the years there have been many numerical studies and calculations associated with the KP equation cf.  \cite{Kako80, Xp94, Kle07, Abl11b, Kao10}.

Our  interest is to study DSW systems which have a discontinuity across a parabolic front; this is analogous  to the well known Riemann or shock tube problem in classical shock waves. We find that indeed there are DSWs  generated across the shock front. To our knowledge this is the first time the nondecaying $2+1$ analogue of a Riemann problem for KP and 2DBO is analyzed in detail.

The reduction discussed here, which we term parabolic front tracking,   was employed as self-similar reductions \cite{Cat90, Sio94} in the analysis of the Khokhlov-Zabolatskaya (KZ) equation \cite{Zab69} (see also \cite{Lin48}). Indeed the KP/2DBO equation reduces to the KZ equation in the limit of zero dispersion. When viscosity is added to the KZ equation the relevant shock waves  are strongly convergent. Our study of the KP/2DBO DSWs  requires
critical use of Whitham modulation theory, which is necessary due to the weak convergence of the DSWs.

This paper is organized as follows. In Section 2 we reduce the KP and 2DBO equations to the cKdV and cBO equations along a parabolic front. In Section 3 we employ perturbation theory \cite{Luk66} to find the conservation laws associated with Whitham theory for the KdV and cKdV equations. We then transform the Whitham modulation equations employing Riemann-type variables; the resulting Whitham system is not  immediately diagonalizable. We solve the 1+1 Whitham system associated with the KdV and cKdV equations numerically and reconstruct the DSW solutions of KdV and cKdV. We then compare these results  with direct numerical  simulations of KdV and cKdV and show that, apart from an unimportant phase they are in excellent agreement. We also note that the Whitham equations for cKdV exhibit a small discontinuity. This discontinuity would be resolved by taking into account higher order terms (see \cite{Abl70}), but doing so is outside the scope of this paper. In Section 4 the BO and cBO equations are analyzed in the same way as KdV and cKdV are analyzed in Section 3. In Section 5 we compare the 1+1 results for cKdV/cBO and the 2+1 results for KP/2DBO by direct numerical simulations. After accounting for an unimportant mean term  we again find excellent agreement; animations are also included as part of our $2+1$ description. We conclude in Section 6.

\section{Reduction of KP, 2DBO equations  to cKdV, cBO equations}

In this section, we examine DSW propagation associated with  two different $(2+1)$ dimensional nonlinear partial differential equations (PDEs). One is  the Kadomtsev-Petviashvili (KP) equation
\begin{equation}
\label{kp}
\left(u_t+uu_x+\epsilon^2 u_{xxx}\right)_{x}+\lambda u_{yy}=0
\end{equation}
where $\epsilon, \lambda$ are constant. This equation was first derived by Kadomtsev-Petviashvili \cite{KP70}  in the context of plasma physics; subsequently it was derived in water waves \cite{Abl79} where it describes the evolution of weakly nonlinear two dimensional long water waves of small amplitude.
When $|\epsilon| \ll1$ we have weak dispersion. According to the sign of $\lambda$, Eq.~(\ref{kp})  is usually termed  \emph{KP-I} $(-)$ or  \emph{KP-II} $(+)$, respectively. KP-I describes the
dynamics when the surface tension of the water is strong and KP-II describes the dynamics  with weak surface tension.
The other equation we study is
\begin{equation}
\label{kpbo}
\left(u_t+uu_x+\epsilon \mathcal{H}\left(u_{xx}\right)\right)_{x}+ \lambda u_{yy}=0
\end{equation}
where $\mathcal{H}u(x)$ denotes the Hilbert transform:

\begin{equation}
\label{Hilbert}
\mathcal{H}u(x)= \frac{1}{\pi}\mathcal{P}\int_{-\infty}^{\infty} \frac{u(x')}{x'-x}dx'
\end{equation}
and $\mathcal{P}$ denotes the Cauchy principal value. We refer to  Eq.~(\ref{kpbo}) as the 2DBO (Two Dimensional Benjamin-Ono) equation; it is a  two-dimensional extension of the classical BO equation and describes weakly nonlinear long internal waves in fluids of great depth \cite{Abl80}.

The goal in this paper is to enhance understanding of DSWs in multidimensional systems. A general form for  these two equations is
\begin{equation}
\label{geq}
\left(u_t+uu_x+F_i(u)\right)_{x}+\lambda u_{yy}=0;
\end{equation}
when $F_1(u)=\epsilon^2 u_{xxx}$
Eq.~(\ref{geq}) is the KP equation and when $F_2(u)=\epsilon \mathcal{H}\left(u_{xx}\right)$
it is the 2DBO equation. Also, we are interested in a class of  initial conditions for Eq.~(\ref{geq}) which are almost step-like initial data, for example
\begin{equation}
\label{ic}
u(x,y,0)=\frac{1}{2}\left(1+\mu\, \textrm{tanh}\left(K\left(x+\frac{1}{2}P(y,0)\right)\right)\right),
\end{equation}
where $\mu=\pm1, K$ are real constants.

\begin{figure}[]
\centering
\subfigure[]{
\includegraphics[width=0.48\textwidth]{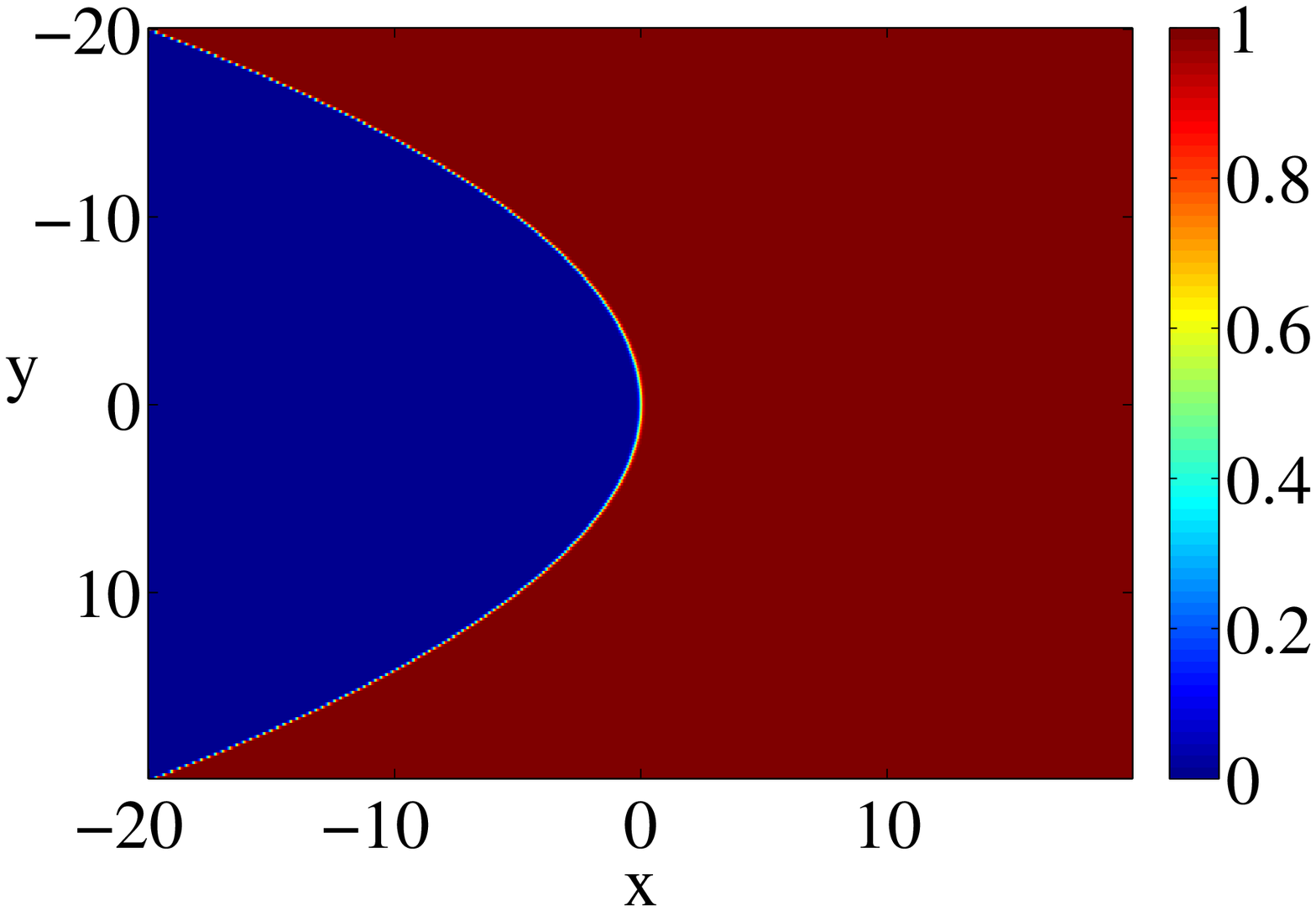}}
\subfigure[]{
\includegraphics[width=0.48\textwidth]{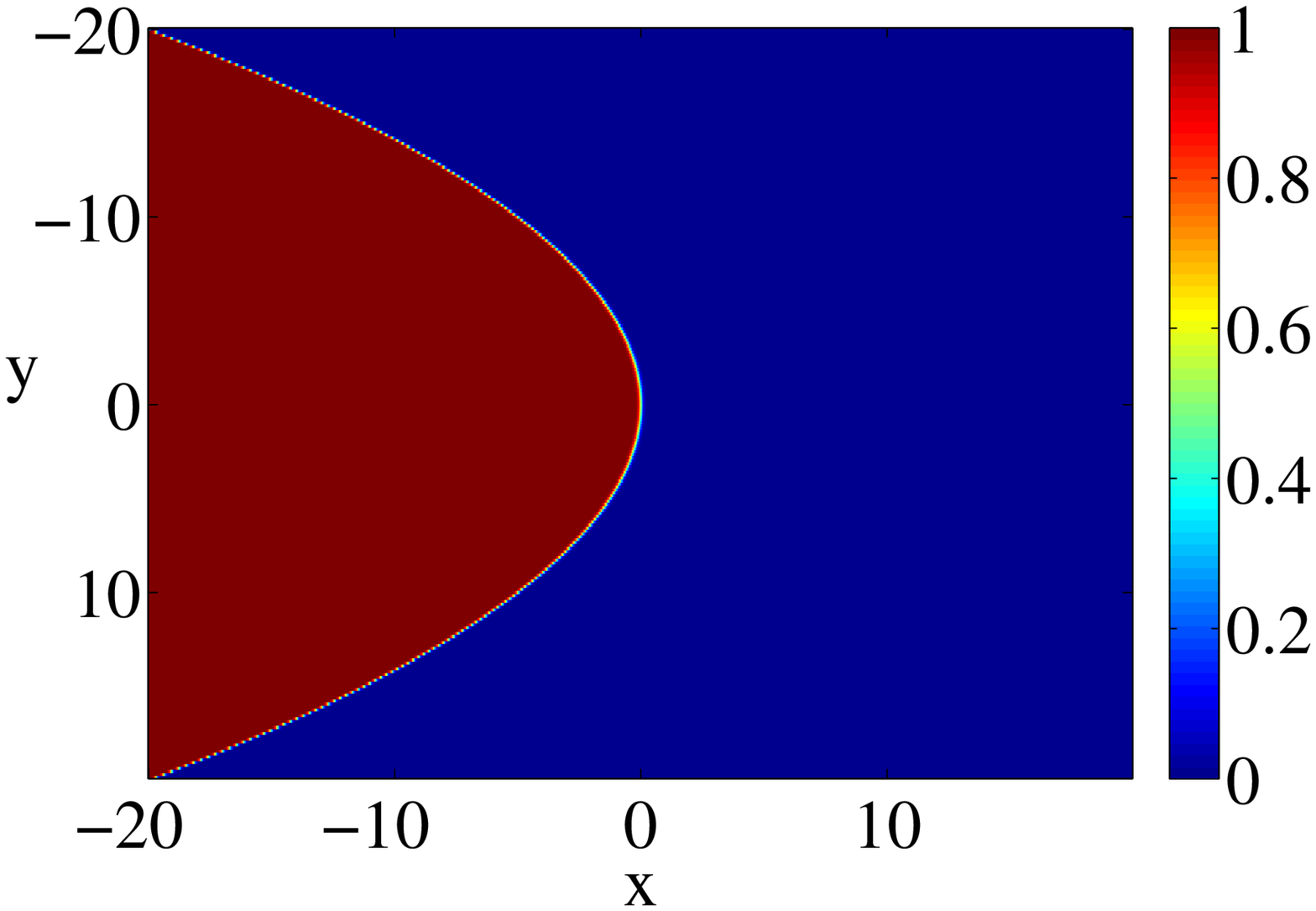}}
\caption{\small Contour plots of initial conditions for $P(y,0)=\tilde{c}y^2$ where $\tilde{c}=0.1$, $K=10$, and (a) $\mu=1$ (non-decreasing in $x$);
(b) $\mu=-1$ (non-increasing in $x$).}
\label{Fig1}
\end{figure}

The above front (\ref{ic}) is a regularized two dimensional extension of the Riemann-type initial condition
\begin{equation}
\label{ric}
u(\eta,0) = \left\{
\begin{array}{lr}
 1, & \eta <0;\\
 0, &  \eta \geq0,
 \end{array}
 \right.
\end{equation}
where $\eta=x+\frac{1}{2}P(y,0)$.
For the special choice of a parabolic front
\begin{equation}\label{eq:Pcy2}
P(y,0)=\tilde{c}y^2,
\end{equation}
where $\tilde{c}$ is a real constant, graphs of the initial conditions with $\mu=1$ ($\mu=-1$) are given in Fig.~\ref{Fig1}.  Note that these initial data increase (decrease) in $x$ across the parabolic front.

We assume that solutions of Eq.~(\ref{geq}) satisfy the following ansatz:
\begin{equation}
\label{anz}
u=f(x+P(y,t)/2,t,y);
\end{equation}
the front is then described by $x+P(y,t)/2=$ constant. We substitute the  ansatz (\ref{anz}) into Eq.~(\ref{geq}) and find
\begin{equation}
\label{rel1}
\left(\frac{1}{2}P_{t}f_{\eta}+f_t+ff_{\eta}+F_i\left(f(\eta)\right)\right)_{\eta}+\lambda\left(\frac{1}{4}\left(P_{y}\right)^2f_{\eta\eta}+\frac{1}{2}P_{yy}f_{\eta }+P_yf_{\eta y}+ f_{yy}\right)=0
\end{equation}
where $\eta=x+P(y,t)/2$. In the above equation $u$ satisfies the following boundary conditions at the infinities:
for non-decreasing type initial conditions
\begin{equation}
\label{bc1}
u\rightarrow0\,\,\,\, \textrm{as} \,\,\, \eta\rightarrow-\infty\,\,\,\,\,\, \textrm{and} \,\,\,\,\,\,u\rightarrow R(t)\,\,\,\, \textrm{as} \,\,\,\,\,\eta\rightarrow\infty
\end{equation}
and for non-increasing type initial conditions
\begin{equation}
\label{bc2}
u\rightarrow R(t)\,\,\,\, \textrm{as} \,\,\, \eta\rightarrow-\infty\,\,\,\,\, \textrm{and} \,\,\,\,\,\,u\rightarrow0\,\,\,\, \textrm{as} \,\,\,\,\,\eta\rightarrow\infty.
\end{equation}
The function $R(t)$ is chosen appropriately; the forms of  $R(t)$ are given later in this section.

Using these boundary conditions and assuming that $P_{yy}$ is independent of $y$, the coefficient of the term $f_{\eta}$ vanishes in Eq.~(\ref{rel1}) and thus $f$ can be taken to be independent of $y$. Then we obtain the following system of equations
\begin{subequations}
\label{eqs}
\begin{equation}
\label{peq}
P_t+\frac{\lambda}{2}\left(P_y\right)^2=0,\,\,\,(front\, shape\, equation:\, FS)
\end{equation}
\begin{equation}
\label{feq}
f_{t}+ff_{\eta}+\frac{\lambda}{2}P_{yy}f+F_i\left(f(\eta)\right)=0.
\end{equation}
\end{subequations}
The FS equation can be transformed into the inviscid Burgers equation,  also sometimes called the Hopf Equation \cite{Hopf50}:
\begin{equation}
\label{veq}
v_t+ \lambda vv_y=0,
\end{equation}
by taking $v=P_y$. This equation is exactly solvable by the method of characteristics. For the parabolic front initial condition (\ref{eq:Pcy2}), we get the following initial condition for Eq.~(\ref{veq}):
\begin{equation}
\label{vic1}
v(y,0)=2\tilde{c}y.
\end{equation}
The solution of the initial value problem (IVP) (\ref{veq}) and (\ref{vic1}) is found as
\begin{equation}
\label{vs1}
v(y,t)=\frac{2\tilde{c}y}{1+2\tilde{c} \lambda t}.
\end{equation}
Thus the front shape $P(y,t)$ is given by
\begin{equation}
\label{vs1}
P(y,t)=\frac{\tilde{c}y^2}{1+2\tilde{c} \lambda t}.
\end{equation}
Consistent with our original assumption, the front curvature $P_{yy}$ is indeed independent of $y$ for all time.

At fixed time $t$, the level curves  of the solution $\eta=x+\frac{P(y,t)}{2} = \eta_0$,  $\eta_0$ constant, are along the following curves in the $(x,y)$-plane:
\begin{equation}
\label{center}
x= -\frac{\tilde{c}y^2}{2(1+2\lambda \tilde{c}t)} +\eta_0= C(t)y^2+\eta_0, ~~C(t)=- \frac{\tilde{c}}{2(1+2\lambda \tilde{c}t)}.
\end{equation}
This shows that when $\tilde{c} \lambda >0$ the curvature of the initial parabolic front decreases in the positive $t$ direction, or equivalently the parabolic front `flattens' when $t$ increases as indicated by $C(t)$. However, the curvature blows up in the negative $t$ direction at a critical time $t_c=-1/(2 \lambda \tilde{c})$. In the following we will take $\tilde{c} \lambda>0$ and only be concerned with $t>0$ to avoid blowup. Evidently this reduction is not valid for all time, but this is common with self-similar reductions as they frequently describe asymptotic phenomena.

In summary, we have shown that for the parabolic front initial condition (\ref{eq:Pcy2}), by  using the ansatz (\ref{anz}), the $(2+1)$ dimensional PDE (\ref{geq}) can be exactly reduced to a $(1+1)$ dimensional PDE (\ref{feq}) with variable coefficients.This $1+1$ dimensional PDE is either the cylindrical KdV (cKdV) equation
\begin{equation}
\label{cKdV}
f_{t}+ff_{\eta}+\frac{\lambda \tilde{c}}{1+2\lambda \tilde{c}t}f+\epsilon^2 f_{\eta\eta\eta}=0
\end{equation}
or the cylindrical BO (cBO) equation
\begin{equation}
\label{cBO}
f_{t}+ff_{\eta}+\frac{\lambda \tilde{c}}{1+2\lambda \tilde{c}t}f+\epsilon\mathcal{H}\left(f_{\eta\eta}\right)=0
\end{equation}
depending on the choice of $F_i(u)$ in Eq.~(\ref{geq}). We reiterate that  the solution (\ref{vs1}) shows that the term $P_{yy}$ in (\ref{feq}) is independent of $y$. In the next two sections, we examine the DSW solutions of Eqs.~(\ref{cKdV}) and (\ref{cBO}) with (non-increasing) initial data  such as Eq.~(\ref{ric}).

Later we will denote $t_0=\frac{1}{\lambda \tilde{c}}$ so the term $\frac{\lambda \tilde{c}}{1+2 \lambda \tilde{c}t}$ becomes $\frac{1}{(2t+t_0)}$. Also, we will consider only $\lambda=1$ (i.e,. KPII, 2DBOII); the other sign (KPI, 2DBOI) can be obtained by changing $\tilde{c}$ to $-\tilde{c}$, i.e. changing the direction of the parabolic front.

In order to determine the boundary conditions associated with Eq.~(\ref{cKdV}-\ref{cBO}) at infinity, we neglect $\eta$ dependent terms and then solve the remaining ODE with the corresponding initial condition (\ref{ric}). The solution of this ODE with the initial condition $R(0)=1$ determines the function $R(t)$ in the boundary conditions (\ref{bc1}) and (\ref{bc2}) as
\begin{equation}
\label{rr}
R(t)=\frac{1}{\sqrt{1+2\tilde{c}t}}.
\end{equation}

We  note that when the parabolic front initial condition (\ref{eq:Pcy2}) is not satisfied, the term $P_{yy}$ in (\ref{feq}) is no longer independent of $y$, and so Eq.~(\ref{feq}) cannot describe the front evolution self-consistently. In this case, taking the FS equation (\ref{peq}) to be the same, Eq.~(\ref{feq}) then changes to
\begin{equation}
\label{feqn}
\left(f_t+\frac{1}{2}P_{yy}f+ff_{\eta}+F_i\left(f(\eta)\right)\right)_{\eta}+ \lambda (P_yf_{\eta y}+{f_{yy}} )= 0.
\end{equation}
This new equation is a $(2+1)$ PDE which at this point is more complicated than the original equation (\ref{geq}). Hence the ansatz (\ref{anz}) for general initial conditions does not lead to cKdV or cBO  since there are additional terms which depend on $f_y, f_{yy}$. However for a class of initial conditions whose evolution in time  satisfies the condition
\begin{equation}
\label{intr}
|L(f;P)| \ll 1
\end{equation}
where $L(f;P)= P_y f_{\eta y} + f_{yy}$, the system (\ref{eqs}) still approximately describes solutions of the equation (\ref{geq}) perhaps for a finite time and/or a subdomain in space. The condition (\ref{intr}) means the additional terms in Eq.~(\ref{feqn}) can be neglected and Eq.~(\ref{feqn}) turns into Eq.~(\ref{feq}) (with $\lambda=1$).

\bigskip

\section{Dispersive shock waves in the  KdV and cKdV equations}

In this section we will investigate the DSW solutions associated with both the KdV and cKdV equations via Whitham modulation theory. We will compare Whitham theory and direct numerical simulations and show that they agree well. Since the KdV equation may be viewed as a special case of the cKdV equation  (\ref{cKdV}) when $\tilde{c}=0$, we will develop the Whitham approach only for cKdV.

From Whitham modulation theory we find three conservation laws. Then, we transform these three conservation laws into a system of quasilinear first order PDEs
by using convenient Riemann type variables; this was first introduced/derived by Whitham for the KdV equation \cite{Whi65}. For KdV this system  can be diagonalized and solved exactly. The system for the cKdV equation cannot be  immediately diagonalized. Numerically we show that the system has solutions which demonstrate the DSW structure of cKdV, unlike KdV, decays in time.

We will use a method of multiple scales originally employed by Luke \cite{Luk66} in the study of Whitham type systems associated with a nonlinear Klein-Gordon equation. For the cKdV equation the leading order equation has a Jacobian elliptic function solution, i.e. the cnoidal wave solution, where the parameters are slowly varying; there are three independent parameters. The leading order problem introduces the rapidly varying phase which requires a compatibility condition which is often termed conservation of waves.  The next order in the perturbation method has two secularity conditions; these together with conservation of waves give three conservation laws.

\subsection{Whitham modulation equations for KdV/cKdV -- conservation laws}

In what follows we develop the slowly varying Whitham modulation equations.
We assume $f=f(\theta, \eta,t;\epsilon)$ where $\theta$ is rapidly varying and defined from

\begin{equation}
\label{phase}
\theta_{\eta}=\frac{k(\eta,t)}{\epsilon},\,\,\,\,\,\,\theta_{t}=-\frac{\omega(\eta,t)}{\epsilon}=-\frac{kV}{\epsilon}
\end{equation}
where $k$, $\omega$ and $V$ are the wave number, frequency and phase velocity, respectively. This definition gives us the compatibility condition $\left(\theta_{\eta}\right)_t=\left(\theta_t\right)_{\eta}$ (conservation of waves) as
\begin{equation}
\label{comp}
k_t+\left(kV\right)_{\eta}=0.
\end{equation}
This is the first conservation law.

With these rapid and slow variables we transform Eq.~(\ref{cKdV}) using
\begin{equation}
\label{vartransf}
\frac{\partial}{\partial \eta} \rightarrow \frac{k}{\epsilon} \frac{\partial}{\partial \theta} + \frac{\partial}{\partial \eta}, ~~\frac{\partial}{\partial t} \rightarrow -\frac{\omega}{\epsilon}  \frac{\partial}{\partial \theta} + \frac{\partial}{\partial t}~
\end{equation}
we have
$$[(-\frac{\omega}{\epsilon} \frac{\partial}{\partial \theta} +  \frac{\partial}{\partial t}) + \epsilon^2 ( \frac{k}{\epsilon}  \frac{\partial}{\partial \theta} + \frac{\partial}{\partial \eta})^3]f+ f ( \frac{k}{\epsilon}  \frac{\partial}{\partial \theta}+ \frac{\partial}{\partial \eta}) f
+ \frac{\lambda \tilde{c}}{1+2 \lambda  \tilde{c} t}f =0$$
or
\begin{equation}
\label{orderkdv}
\begin{aligned}
&\frac{1}{\epsilon}\left(-\omega\frac{\partial f}{\partial \theta}+k f \frac{\partial f}{\partial \theta}+k^3\frac{\partial^{3} f}{\partial \theta^{3}}\right)+\left(\frac{\partial f}{\partial t}+f\frac{\partial f}{\partial \eta}+3kk_{\eta}\frac{\partial^{2} f}{\partial \theta^{2}}+3k^2\frac{\partial^{3} f}{\partial \theta^{2} \partial \eta}+\frac{\lambda \tilde{c}}{1+2 \lambda \tilde{c} t}f\right)\\
&+\epsilon\left(k_{\eta\eta}\frac{\partial f}{\partial \theta}+k_{\eta}\frac{\partial^{2} f}{\partial \theta \partial \eta}+k\frac{\partial^{3} f}{\partial \theta \partial \eta^{2}}+2k\frac{\partial^{3} f}{\partial \theta^{2} \partial \eta}\right)+\epsilon^{2}\frac{\partial^{3} f}{\partial \eta^{3}}=0.
\end{aligned}
\end{equation}
Then we expand $f$ in powers of $\epsilon$  as
\begin{equation}
\label{exp2}
f\left(\theta,\eta,t\right)=f_0\left(\theta,\eta,t\right)+\epsilon f_1\left(\theta,\eta,t\right)+... .
\end{equation}
Grouping the terms in like powers of $\epsilon$ gives  leading and higher order perturbation  equations; we only consider the first two orders here. The $\mathcal{O}\left(\frac{1}{\epsilon}\right)$ equation is
\begin{equation}
\label{lo2}
-\omega f_{0,\theta}+k f_{0}f_{0,\theta}+k^{3}f_{0,\theta\theta\theta}=0;
\end{equation}
the $\mathcal{O}(1)$ equation is
\begin{equation}
\label{sop}
\mathcal{L}f_{1}\equiv-\omega f_{1,\theta}+k\left(f_{0}f_{1}\right)_{\theta}+k^{3}f_{1,\theta\theta\theta}=G,
\end{equation}
where
\begin{equation}
\label{gg}
G\equiv-\left(f_{0,t}+f_{0}f_{0,\eta}+3kk_{\eta}f_{0,\theta\theta}
+3k^{2}f_{0,\theta\theta\eta}+\frac{f_0}{2t+t_0}\right).
\end{equation}
We can proceed to higher order terms, but doing so is outside the scope of this paper.

The solution of (\ref{lo2}) is
\begin{equation}
\label{los}
f_0\left(\theta,\eta,t\right)=a\left(\eta,t\right)+b\left(\eta,t\right)\textrm{cn}^2\left[2\left(\theta-\theta_{0}\right)K,m\left(\eta,t\right)\right]
\end{equation}
where $K\equiv K\left(m\left(\eta,t\right)\right)$ is the complete elliptic integral of the first kind and
\begin{equation}\label{eq:k2a}
 k^2=\frac{b}{48m^{2}K^2},\,\,\,\,\,\,a=V+\frac{b}{3m^2}-\frac{2b}{3};
\end{equation}
and recall that $V= \frac{\omega}{k}. $ For the purposes of this paper we consider $\theta_0$ to be a constant. At this point we have three independent parameters: $b,V,m$.

We rewrite the conservation law (\ref{comp}) by using the above formulae as
\begin{equation}
\label{clw1}
\frac{\partial}{\partial t}\left(\frac{1}{4\sqrt{3}K}\sqrt{\frac{b}{m^2}}\right)+\frac{\partial}{\partial \eta}\left(\frac{V}{4\sqrt{3}K}\sqrt{\frac{b}{m^2}}\right) =0.
\end{equation}
When we use the solution (\ref{los}) in (\ref{sop}),  secular terms can occur, i.e. terms that grow arbitrarily large with respect to $\theta$.
Let $w$ denote solutions of the adjoint problem to $\mathcal{L}u=0$, i.e.,
\begin{equation}
\label{adj1}
\mathcal{L}^{A}w=0,\,\,\,\,\,\mathcal{L}^{A}=\omega \partial_\theta-k f_{0}\partial_\theta-k^{3}\partial_{\theta\theta\theta}.
\end{equation}
To eliminate the secular terms,
we use the following relation obtained from Eq.~(\ref{sop})
\begin{equation}
\label{adj2}
\int_{0}^{1}[w\mathcal{L}f_{1}-f_{1}\mathcal{L}^{A}w]d\theta=\int_{0}^{1}wGd\theta.
\end{equation}
The adjoint problem (\ref{adj1}) has two linearly independent solutions $w=1$ and $w=f_0$, the latter following from Eq.~(\ref{lo2}). We substitute these into Eq.~(\ref{adj2}), enforce the periodicity of $f_0\left(\theta,\eta,t\right)$ in $\theta$ and obtain the following secularity conditions
\begin{equation}
\label{secs}
\int_{0}^{1}Gd\theta=0,\,\,\,\,\,\textrm{and}\,\,\,\,\int_{0}^{1}f_{0}Gd\theta=0.
\end{equation}
Using (cf. \cite{Abl13})
\begin{equation}
\label{int1}
\int_{0}^{1}\frac{\partial^{i}f_0}{\partial \theta^i}d\theta=0, \,\,\,\,\,\int_{0}^{1}f_0\frac{\partial^{j}f_0}{\partial \theta^j}d\theta=0
\end{equation}
for $i=1,2,...$ and $j=1,3,...$, and
\begin{equation}
\label{int2}
\int_{0}^{1}f_{0}f_{0,\theta\theta}d\theta=-\int_{0}^{1}f_{0,\theta}^{2}d\theta,
\end{equation}
we get from the first secularity condition in (\ref{secs})
\begin{equation}
\label{iconl1}
\frac{\partial}{\partial t}\int_{0}^{1}f_{0}d\theta+\frac{\partial}{\partial \eta}\left(\frac{1}{2}\int_{0}^{1}f_{0}^{2}d\theta\right)
+\frac{1}{2t+t_0}\int_{0}^{1}f_{0}d\theta=0
\end{equation}
and the second secularity condition in (\ref{secs})
\begin{equation}
\label{iconl2}
\frac{\partial}{\partial t}\int_{0}^{1}f_{0}^{2}d\theta+\frac{\partial}{\partial \eta}\left(\frac{2}{3}\int_{0}^{1}f_{0}^{3}d\theta
-3k^2\int_{0}^{1}f_{0,\theta}^{2}d\theta\right)+\frac{2}{2t+t_0}\int_{0}^{1}f_{0}^{2}d\theta=0.
\end{equation}
We can use Eq.~(\ref{los}) to rewrite the conservation laws (\ref{iconl1}) and (\ref{iconl2}) in terms of $m$, $V$ and $b/m^2$. From the properties of the elliptic functions \cite{byrd71}, we find
\begin{equation}
\label{ident}
\begin{aligned}
&\int_{0}^{1}f_{0}d\theta=V+\frac{b}{3m^2}\left(\frac{3E}{K}+m^{2}-2\right),\\
&\int_{0}^{1}f_{0}^{2}d\theta=V^{2}+2V\frac{b}{3m^2}\left(\frac{3E}{K}+m^{2}-2\right)+\left(\frac{b}{3m^2}\right)^{2}\left(m^{4}-m^{2}+1\right),\\
&\int_{0}^{1}f_{0}^{3}d\theta=V^{3}+V^2\frac{b}{m^2}\left(\frac{3E}{K}+m^{2}-2\right)+\frac{V}{3}\left(\frac{b}{m^2}\right)^{2}\left(m^{4}-m^{2}+1\right)\\
&+\frac{1}{5}\left(\frac{b}{m^2}\right)^{3}\left[\frac{E}{K}\left(m^{4}-m^{2}+1\right)+\frac{1}{27}\left(5m^{6}-21m^{4}+33m^{2}-22\right)\right],\\
&k^{2}\int_{0}^{1}f_{0,\theta}^{2}d\theta=\frac{1}{45}\left(\frac{b}{m}\right)^{3}\left[\frac{2E}{K}\left(m^{4}-m^{2}+1\right)-\left(m^{4}-3m^{2}+2\right)\right],
\end{aligned}
\end{equation}
where $K\equiv K(m)$ and $E\equiv E(m)$ are the complete elliptic integrals of the first and second kind.

Using the formulae (\ref{ident}) in (\ref{iconl1}) and (\ref{iconl2}) the following conservation laws are obtained
\begin{equation}
\label{cclw2}
\begin{aligned}
&\frac{\partial}{\partial t}\left[V+\frac{b}{3m^2}\left(\frac{3E}{K}+m^{2}-2\right)\right]\\
&+\frac{1}{2}\frac{\partial}{\partial \eta}\left[V^{2}+2V\frac{b}{3m^2}\left(\frac{3E}{K}+m^{2}-2\right)+\left(\frac{b}{3m^2}\right)^{2}\left(m^{4}-m^{2}+1\right)\right]\\
&+\frac{1}{2t+t_0}\left[V+\frac{b}{3m^2}\left(\frac{3E}{K}+m^{2}-2\right)\right]=0
\end{aligned}
\end{equation}
and
\begin{equation}
\label{cclw3}
\begin{aligned}
&\frac{\partial}{\partial t}\left[V^{2}+2V\frac{b}{3m^2}\left(\frac{3E}{K}+m^{2}-2\right)+\left(\frac{b}{3m^2}\right)^{2}\left(m^{4}-m^{2}+1\right)\right]\\
&+\frac{\partial}{\partial \eta}\Bigg[\frac{2V^3}{3}+\frac{2V^2b}{3m^2}\left(\frac{3E}{K}+m^{2}-2\right)+\frac{2V}{9}\left(\frac{b}{m^2}\right)^{2}\left(m^{4}-m^{2}+1\right)\\
&+\frac{1}{81}\left(\frac{b}{m^2}\right)^{3}\left(2m^{2}-1\right)\left(m^{2}+1\right)\left(m^{2}-2\right)\Bigg]\\
&+\frac{2}{2t+t_0}\Bigg[V^{2}+2V\frac{b}{3m^2}\left(\frac{3E}{K}+m^{2}-2\right)+\left(\frac{b}{3m^2}\right)^{2}\left(m^{4}-m^{2}+1\right)\Bigg]=0.
\end{aligned}
\end{equation}
Equations (\ref{clw1}), (\ref{cclw2}) and (\ref{cclw3}) are the three conservation laws which determine $b$, $m$ and $V$.

\subsection{Whitham modulation equations for KdV/cKdV -- Riemann type variables}
We transform these conservation laws by making a suitable change of variables. In particular we parametrize $b,m,V$  in terms of  the following Riemann type variables $r_1,r_2,r_3$:
\begin{equation}
\label{rvar}
b=2(r_{2}-r_{1}),\,\,\,\,m^2=\frac{r_{2}-r_{1}}{r_{3}-r_{1}},\,\,\,V=\frac{1}{3}\left(r_{1}+r_{2}+r_3\right),\,\,\,r_{1}\leq r_{2}\leq r_{3}
\end{equation}
and it follows from Eq.~(\ref{eq:k2a}) that $a=r_{3}-r_{2}+r_{1}$.

This leads to an equation of the form
\begin{equation}
\label{diagA}
\sum_{j=1}^3 (A_{i,j}\frac{\partial r_j}{\partial t}+B_{i,j}\frac{\partial r_j}{\partial \eta}+ C_{i,j})=0,\,\,\,\,\,\,\,\,\,i=1,2,3,
\end{equation}
where $A_{i,j},B_{i,j}$ are functions of $r_i, i=1,2,3$ and $C_{i,j}$ is a function of $r_i, i=1,2,3$ and $t$. Multiplying by the inverse of the $A$ matrix we
can simplify Eq.~(\ref{diagA}) into the following quasilinear PDE system
\begin{equation}
\label{diag}
\frac{\partial r_i}{\partial t}+v_i\left(r_1,r_2,r_3\right)\frac{\partial r_i}{\partial \eta}+\frac{h_i\left(r_1,r_2,r_3\right)}{2t+t_0}=0,\,\,\,\,\,\,\,\,\,i=1,2,3,
\end{equation}
where
\begin{equation}
\begin{aligned}
&v_1=\frac{1}{3}\left(r_{1}+r_{2}+r_3\right)-\frac{2}{3}\left(r_2-r_1\right)\frac{K(m)}{K(m)-E(m)},\\
&v_2=\frac{1}{3}\left(r_{1}+r_{2}+r_3\right)-\frac{2}{3}\left(r_2-r_1\right)\left(1-m^2\right)\frac{K(m)}{E(m)-\left(1-m^2\right)K(m)},\\
&v_3=\frac{1}{3}\left(r_{1}+r_{2}+r_3\right)+\frac{2}{3}\left(r_3-r_1\right)\frac{K(m)}{E(m)},\\
&h_1=\frac{(5E(m)-3K(m))r_{1}-(E(m)+K(m))r_{2}+(K(m)-E(m))r_{3}}{3(E(m)-K(m))},\\
&h_2=\frac{E(m)\left(r_3-r_1\right)\left(r_1-5r_2+r_3\right)-K(m)\left(r_2-r_3\right)\left(r_1+3r_2-r_3\right)}{3\left[E(m)r_1-K(m)r_2+(K(m)-E(m))r_3\right]},\\
&h_3=\frac{(2K(m)-E(m))r_2+(5E(m)-2K(m))r_3-E(m)r_1}{3E(m)}.
\end{aligned}
\end{equation}
The $r_i$'s are called Riemann variables. From Eq.~(\ref{rvar})  the solution of the leading order problem (\ref{los}) is reconstructed  from the the $r_i$'s as
\begin{equation}
\label{soll}
f_0\left(\theta,\eta,t\right)=r_1-r_2+r_3+2\left(r_2-r_1\right)\textrm{cn}^{2}\left[2K\left(\theta-\theta_0\right),m\right].
\end{equation}
The rapid phase $\theta$ is determined by integrating (\ref{phase})
\begin{equation}
\label{iphase}
\theta\left(\eta,t\right)=\int_{-\infty}^{\eta}\frac{k(x^{'},t)}{\epsilon}dx^{'}-\int_{0}^{t}\frac{k(\eta,t^{'})V(\eta,t^{'})}{\epsilon}dt^{'}
\end{equation}
We also note that there is a free constant $\theta_0$ in Eq.~(\ref{soll}) which we determine by comparison with direct numerical simulations.

\begin{figure}[]
\centering
\includegraphics[width=0.48\textwidth]{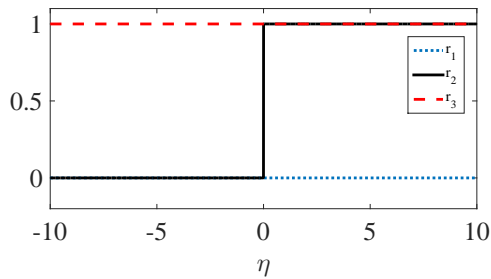}
\caption{\small Initial values (\ref{wic}) for Riemann variables $r_1$, $r_2$ and $r_3$.}
\label{fig2}
\end{figure}

The initial values of the Riemann variables of  the reduced diagonal Whitham system (\ref{diag}) are given below (see Fig.~\ref{fig2})
\begin{equation}
\label{wic}
r_1(\eta,0)=0,\,\,\,\,\,\,\,\,
r_2(\eta,0) = \left\{
\begin{array}{lr}
 0, &  \eta \leq0;\\
 1, &  \eta >0,
 \end{array}
 \right.\,\,\,\,\,\,\,\,r_3(\eta,0)=1.
\end{equation}

In the absence of cylindrical terms, i.e. $h_i =0$,  Eq.~(\ref{diag}) reduces to a diagonal system that agrees with the well known diagonal Whitham system for the KdV equation first derived by Whitham \cite{Whi65} (see also \cite{Gru74}).
For the KdV equation, the solution of the reduced Whitham system provides the dispersive regularization
for the initial data (\ref{ric}). This regularization  can written in terms of a similarity variable $\xi= \eta/t$ and the system (\ref{diag}) reduces to
\begin{equation}
\label{r2sim}
(v_2(r_2)-\xi) r_2'(\xi)=0
\end{equation}
from which we get the self similar solution $v_2(r_2)-\xi=0$ or by inversion $r_2=r_2(\xi)$.
Thus the system (\ref{diag}) with $h_i=0$ admits an exact rarefaction wave solution in terms of the self-similar variable $\xi=\eta/t$. This rarefaction wave solution leads to the DSW solution for the classical KdV equation
(for additonal details cf. \cite{Hoe06}).

However, the system (\ref{diag}) for cKdV is not diagonal and this property makes finding analytical solutions more difficult. The solutions of nondiagonal quasi-linear systems obtained via Whitham modulation theory were investigated in certain cases. In \cite{Gru87}, stationary solutions
were obtained in the KdV-Burgers equation. In \cite{GA07}, the leading and trailing edge structures of DSW solutions were investigated in a variable coefficient KdV (vKdV) equation.

\subsection{Comparison between numerical simulations of Whitham modulation equations and direct numerical simulations of KdV/cKdV}

Our approach is to study the cKdV equation, and the KdV equation as a special case,  by using numerical methods to solve the (nondiagonal, in general)  Whitham modulation equations (\ref{diag}).  We compare the results to direct numerical simulations of the original $1+1$ cKdV  equation. This allows us  to understand the underlying structure of the DSWs in the cKdV equation. Indeed we find very good agreement between the numerical solutions of the Whitham equations and direct numerical simulations of the cKdV equation. We conclude that Whitham modulation theory provides a good approximation of the DSWs in the cKdV equation. The advantage of computing with the Whitham system is that it gives the structure of the DSWs in terms of $\mathcal{O}(1)$ coefficients, whereas for direct numerical simulations one has small coefficients (due to $\epsilon^2 \ll 1$) which in turn requires more sophistication to solve and longer computing times.

First we find numerical solutions of the diagonal reduced Whitham system associated with the KdV equation and the nondiagonal Whitham system (\ref{diag}) for the cKdV equation; the initial values of Riemann variables are given by Eq.~(\ref{wic}).

\begin{figure}[ht]
\centering
\subfigure[]{
\includegraphics[width=0.48\textwidth]{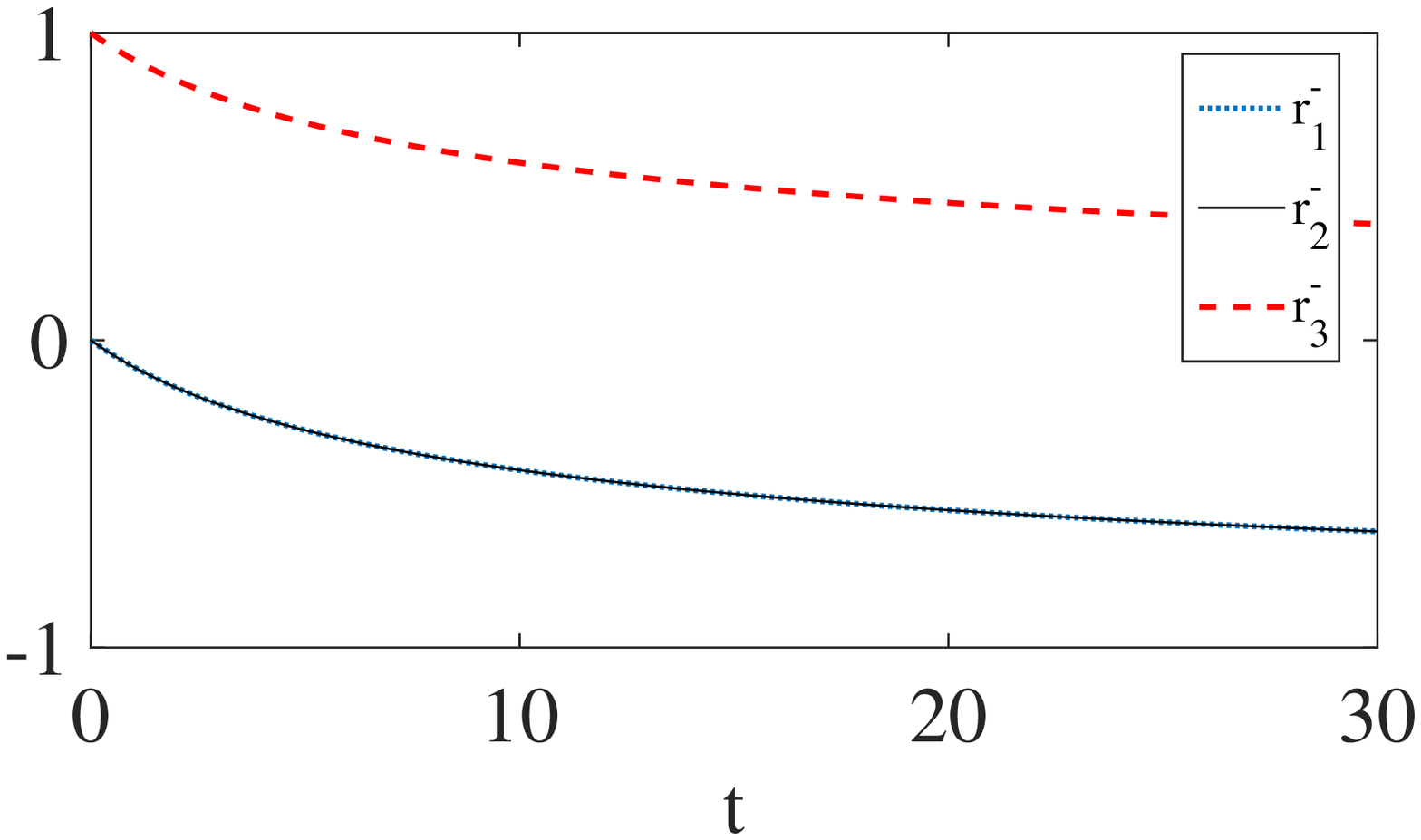}}
\subfigure[]{
\includegraphics[width=0.48\textwidth]{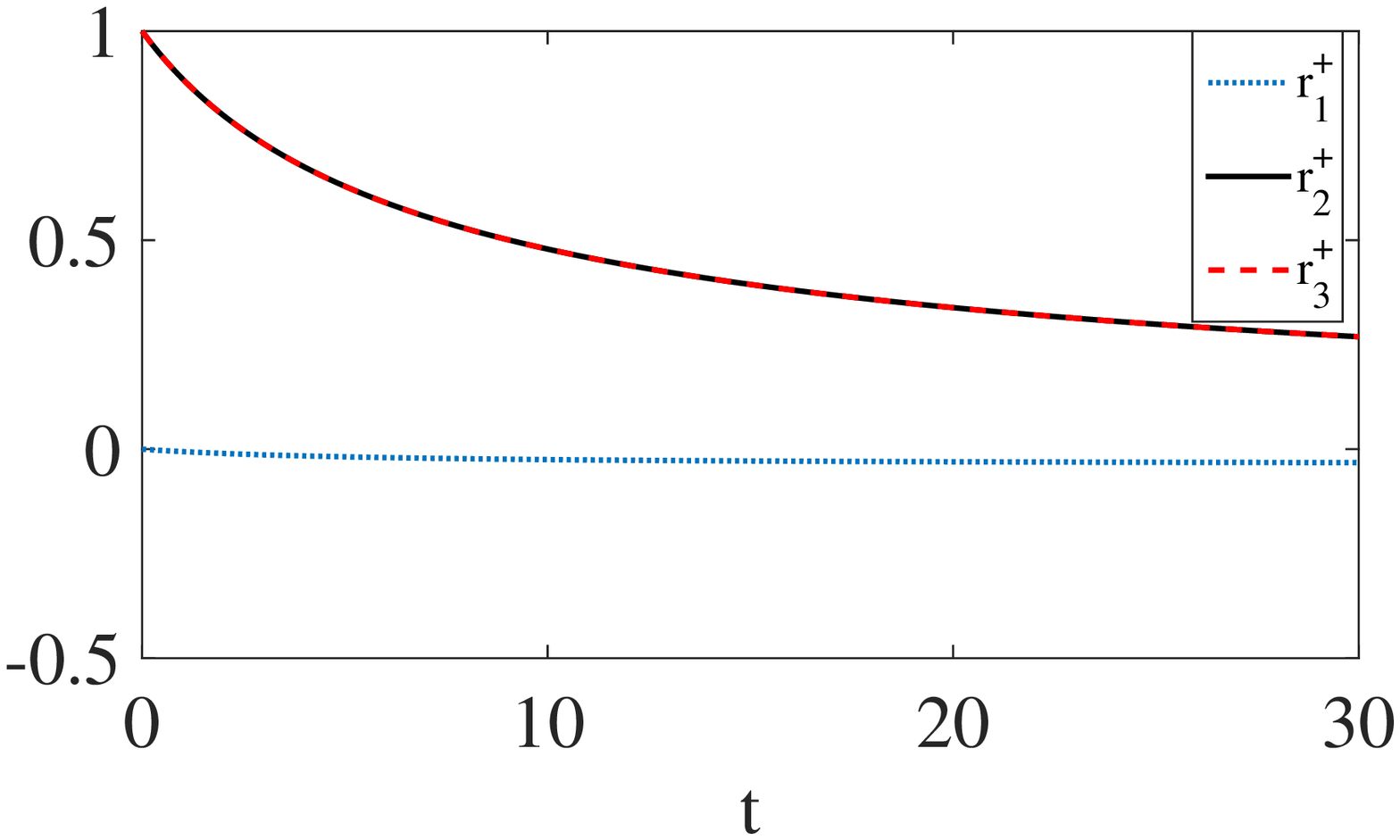}}
\caption{
Evolution of Riemann variables for cKdV case which are obtained by numerical solution of reduced ODE system (\ref{diag}) (a) at the left boundary, (b) at the right boundary.}
\label{fig4a}
\end{figure}

The boundary conditions for Eq.~(\ref{diag})  must be determined before numerical computations can proceed. For the KdV equation and its associated Whitham system, the boundary conditions remain constant at both ends of the domain.
However, the boundary conditions change in time for both the cKdV equation and its associated Whitham system (\ref{diag}).
The boundary conditions  for the cKdV equation are the same as in Eq.~(\ref{bc2}). For the Whitham system (\ref{diag}), we get the corresponding boundary conditions by numerically solving the reduced ODE system obtained from Eq.~(\ref{diag}) by neglecting the spatial variable $\eta$.
This ODE system is solved with the initial conditions (\ref{wic}) at both ends separately by using the ode45 solver of MATLAB$\circledR$. The evolution of the Riemann variables for the cKdV equation  at the boundaries is given in Fig.~\ref{fig4a}. We see that these Riemann variables at the boundaries decay in time.

\begin{figure}[ht]
\centering
\subfigure[]{
\includegraphics[width=0.48\textwidth]{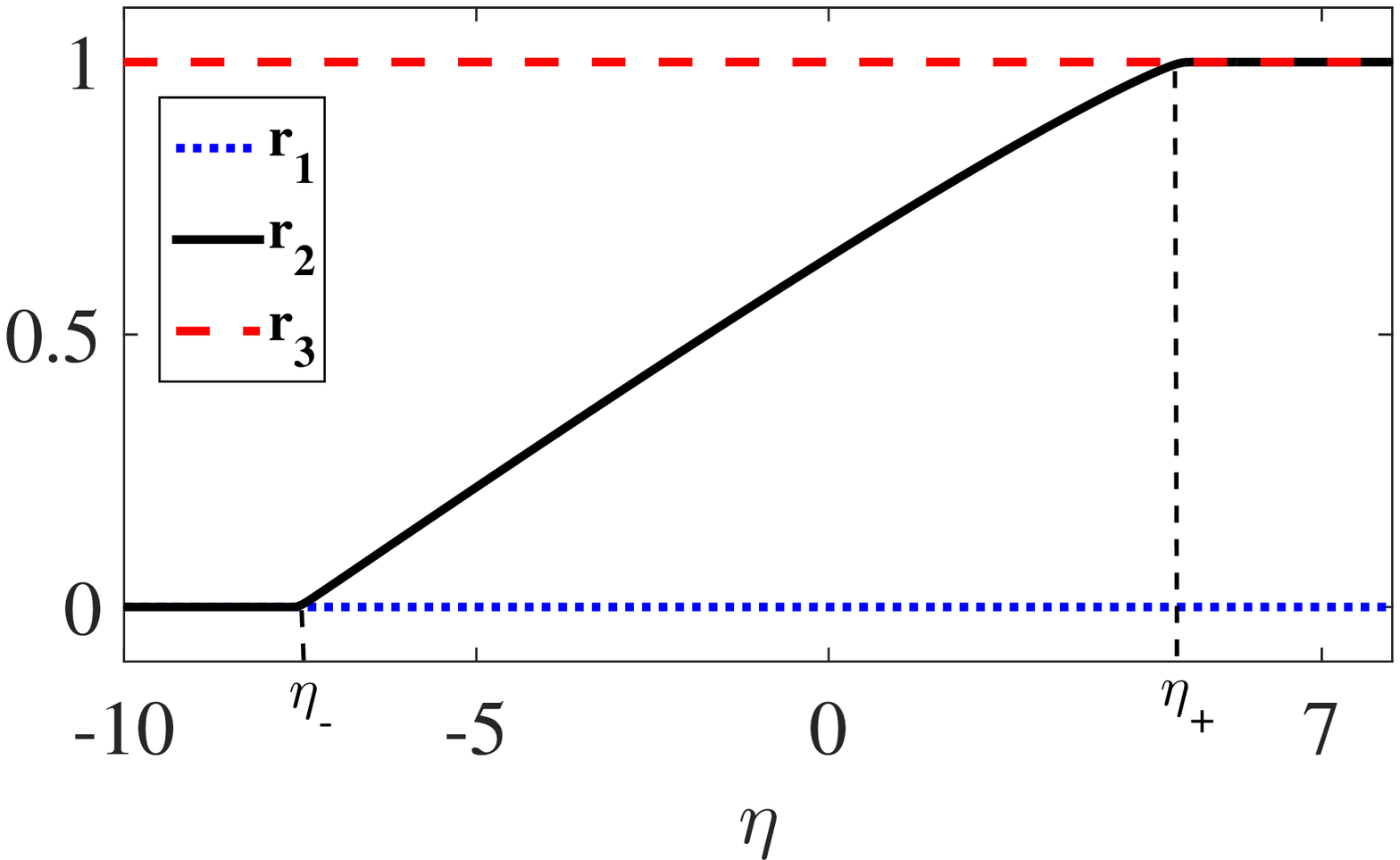}}
\subfigure[]{
\includegraphics[width=0.48\textwidth]{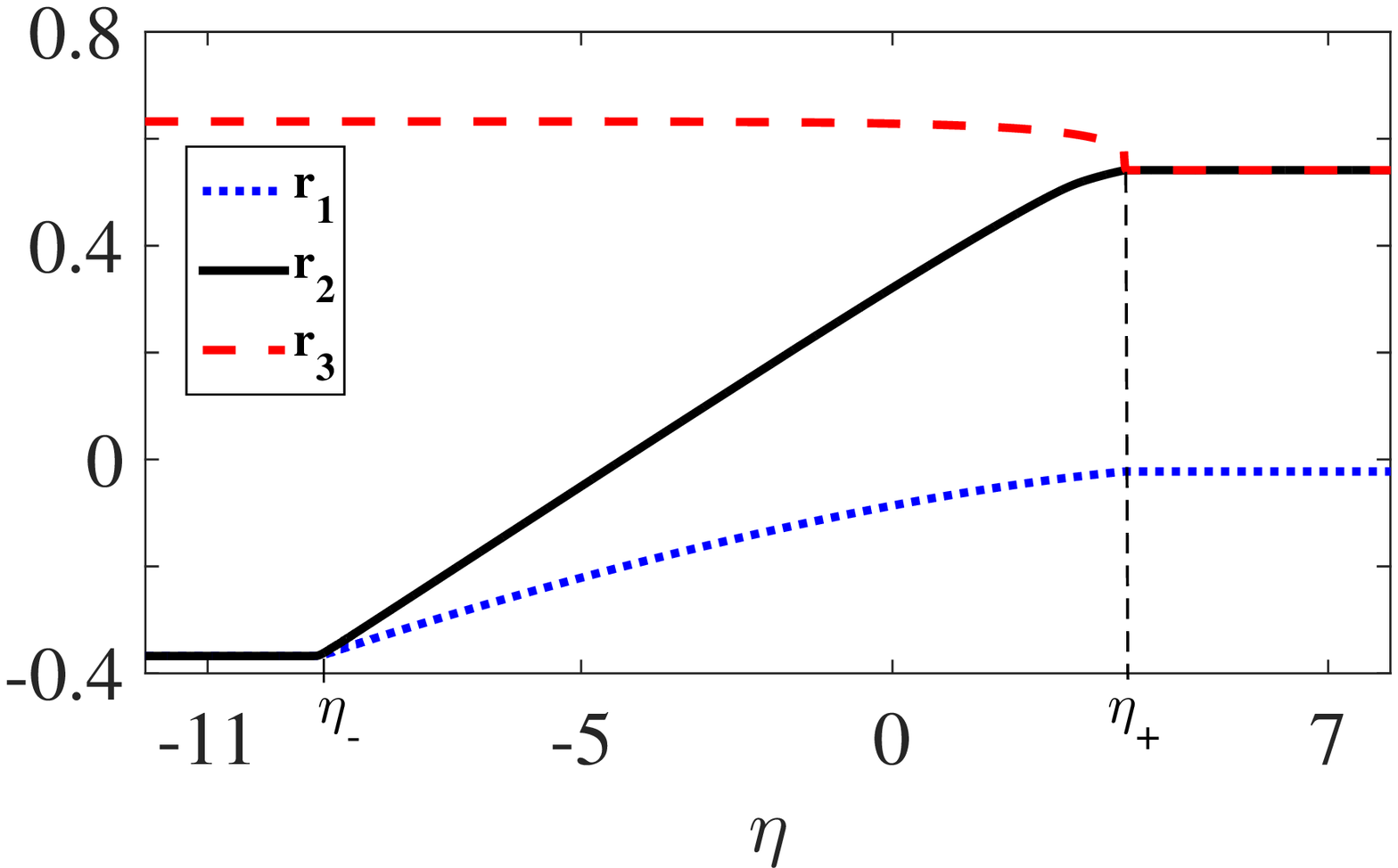}}
\caption{\small Riemann variables at t=7.5 which are found by numerical solutions of Whitham system  (\ref{diag})
(a) for the KdV eq., (b) for the cKdV eq. Here, we take $t_0=10$.}
\label{fig3}
\end{figure}

For the computation of the Whitham system (\ref{diag}) including boundary conditions, we use a first order hyperbolic PDE solver based on MATLAB$\circledR$ by Shampine \cite{Sha05} and choose a two-step variant of the Lax-Wendroff method with a nonlinear filter \cite{Eng89}. In the numerical solutions of Whitham systems, we use $N=2^{12}$ points for the spatial domain $[-30,30]$ with the time step being 0.9 times the spatial step. The resulting numerical solutions are given  in Fig.~\ref{fig3} for both KdV and cKdV.

Interestingly, in the cKdV case the $r_3$ component of the solution of the Whitham system exhibits a small shock-like front in front of the DSW. Direct numerical simulations indicate that this behavior is not significant. In fact  adding higher order terms to the Whitham system is expected to regularize the solutions (see \cite{Abl70}).

\begin{figure}[ht]
\centering
\subfigure[]{
\includegraphics[width=0.48\textwidth]{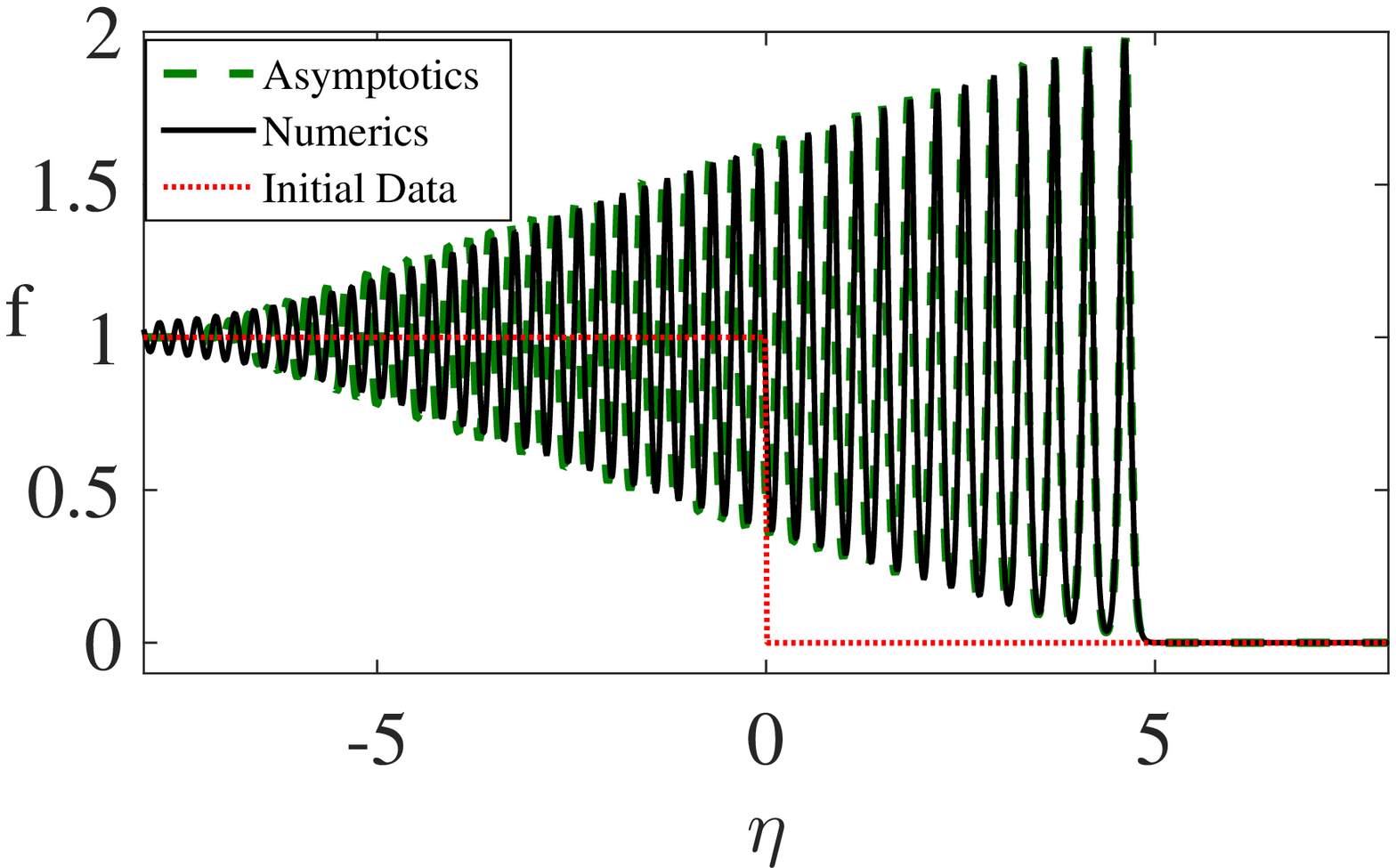}}
\subfigure[]{
\includegraphics[width=0.48\textwidth]{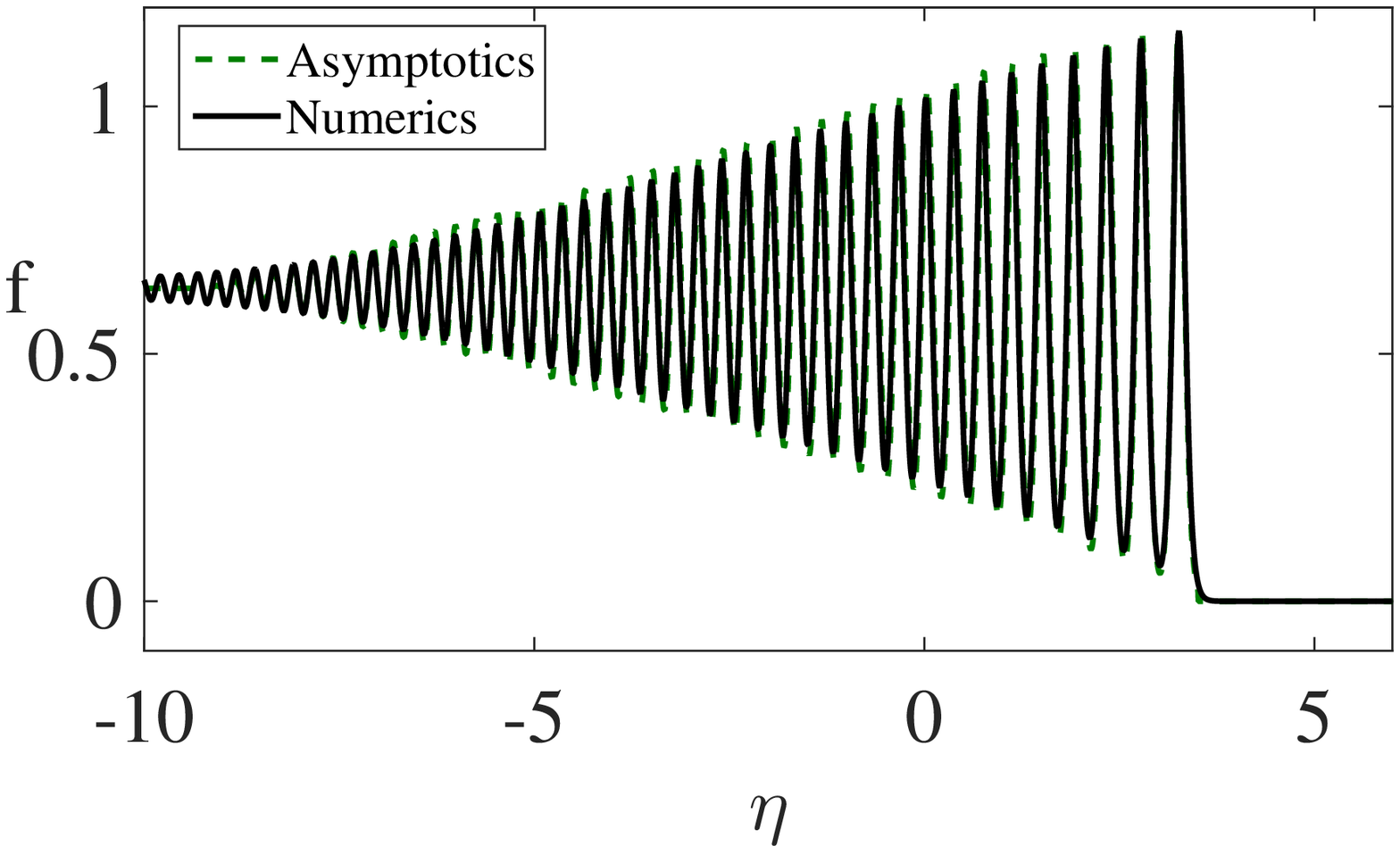}}
\caption{\small Numerical and asymptotic solutions of KdV and cKdV eqs. at t=7.5 with the initial data (\ref{ric}): (a)for KdV eq., (b) for cKdV eq. Here, we take $t_0=10$ and $\epsilon^{2}=10^{-3}$.}
\label{fig5}
\end{figure}

We can reconstruct the corresponding DSWs at any time (e.g. $t=7.5$) for both KdV and cKdV from the Riemann variables $r_i$'s using Eqs.~(\ref{soll}) and (\ref{iphase}).  These are also plotted and compared with direct numerical simulations of KdV/cKdV (discussed below) in  Fig.~\ref{fig5}a and Fig.~\ref{fig5}b. In Fig.~\ref{fig5}b we have chosen the arbitrary phase $\theta_0$ in Eq.~(\ref{soll}) appropriately to agree with the direct numerical simulations.

For the direct numerical simulations of KdV/cKdV, we use a numerical procedure which is useful for problems with fixed boundary conditions. Since the left boundary condition $R(t)$ for cKdV is a function of $t$ we first transform (\ref{cKdV}) by
\begin{equation}
\label{tr1}
f=R(t)\phi
\end{equation}
to the following variable coefficient KdV (vKdV) equation
\begin{equation}
\label{vkd}
\phi_{t}+R(t)\phi\phi_{\eta}+\epsilon^{2}\phi_{\eta\eta\eta}=0,
\end{equation}
where we recall from Eq.~(\ref{rr}) that $R(t)=\frac{\sqrt{t_0}}{\sqrt{2t+t_0}}$ with $t_{0}=\frac{1}{\tilde{c}}$. Equation~(\ref{vkd}) has the left boundary condition fixed at $\phi_{-}=1$, while the right boundary condition $\phi_{+}=0$ stays the same as in the original cKdV equation. In order to solve
{Eq.~(\ref{vkd}) numerically (see also \cite{Abl09, Cox02, Kas05}) we differentiate with respect to $\eta$ and define $\phi_{\eta}=z$ to get
\begin{equation}
\label{zeq}
z_t+R(t)\left(z\phi\right)_{\eta}+\epsilon^2 z_{\eta\eta\eta}=0.
\end{equation}
Transforming to Fourier space gives
\begin{equation}
\label{fzeq}
 \widehat{z_t}= \textbf{L}\widehat{z}+R(t)\textbf{N}\left(\widehat{z},t\right)
\end{equation}
where $\widehat{z}=\mathcal{F}(z)$ is the Fourier transform of $z$, $\textbf{L}\widehat{z}\equiv i\epsilon^2 k^3 \widehat{z}$ and
\begin{equation}
\label{nop}
\textbf{N}\left(\widehat{z},t\right)=-ik\mathcal{F}\left\{\left[\phi_{-}+\int_{-\infty}^{\eta}\mathcal{F}^{-1}\left(\widehat{z}\right)d\eta^{'}\right]
\mathcal{F}^{-1}\left(\widehat{z}\right)\right\}.
\end{equation}
The only difference from the classical KdV case is that for vKdV the nonlinear term $\textbf{N}$ has a time dependent coefficient (see \cite{Abl09}).
To solve the above ODE system in Fourier space we use a modified version of the exponential-time-differencing fourth-order Runge-Kutta (ETDRK4) method \cite{Cox02, Kas05}. For the required spectral accuracy of the ETDRK4 method, the initial condition for $z$ must be smooth and periodic. However, the step initial condition (\ref{ric}) for $u$ or equivalently $f$ leads to $z(\eta,0)=-\delta(\eta)$, where $\delta$ represents the Dirac delta function. Therefore we regularize this initial condition with the analytic function \cite{Abl09}
\begin{equation}
\label{sic}
z\left(\eta,0\right)=-\frac{\tilde{K}}{2}\textrm{sech}^2\left(\tilde{K}\eta\right),
\end{equation}
where $\tilde{K}>0$ is large. Thus Eq.~(\ref{zeq}) can be solved numerically via Eqs.~(\ref{fzeq}-\ref{nop}) on a finite spatial domain $[-L,L]$, where $\mathcal{F}$ represents the discrete Fourier transform and the integration limit $-\infty$ in (\ref{nop}) is replaced by $-L$.

For the ETDRK4 method we take the number of Fourier modes in space to be $N=2^{12}$, the domain size to be $L=30$, and the time step to be $10^{-4}$. The parameters are chosen to be $\tilde{c}^{-1}=t_0=10, \epsilon^2=10^{-3}$ and $\tilde{K}=10$. The numerical results for the KdV equation (with $\tilde{c}=0$) and the cKdV equation are given in Fig.~\ref{fig5}a and Fig.~\ref{fig5}b at $t=7.5$. To provide another view, we also include space-time plots of the direct numerical solutions of KdV and cKdV eqs. are given in Fig.\ref{fig5b}.

\begin{figure}[ht]
\centering
\subfigure[]{
\includegraphics[width=0.48\textwidth]{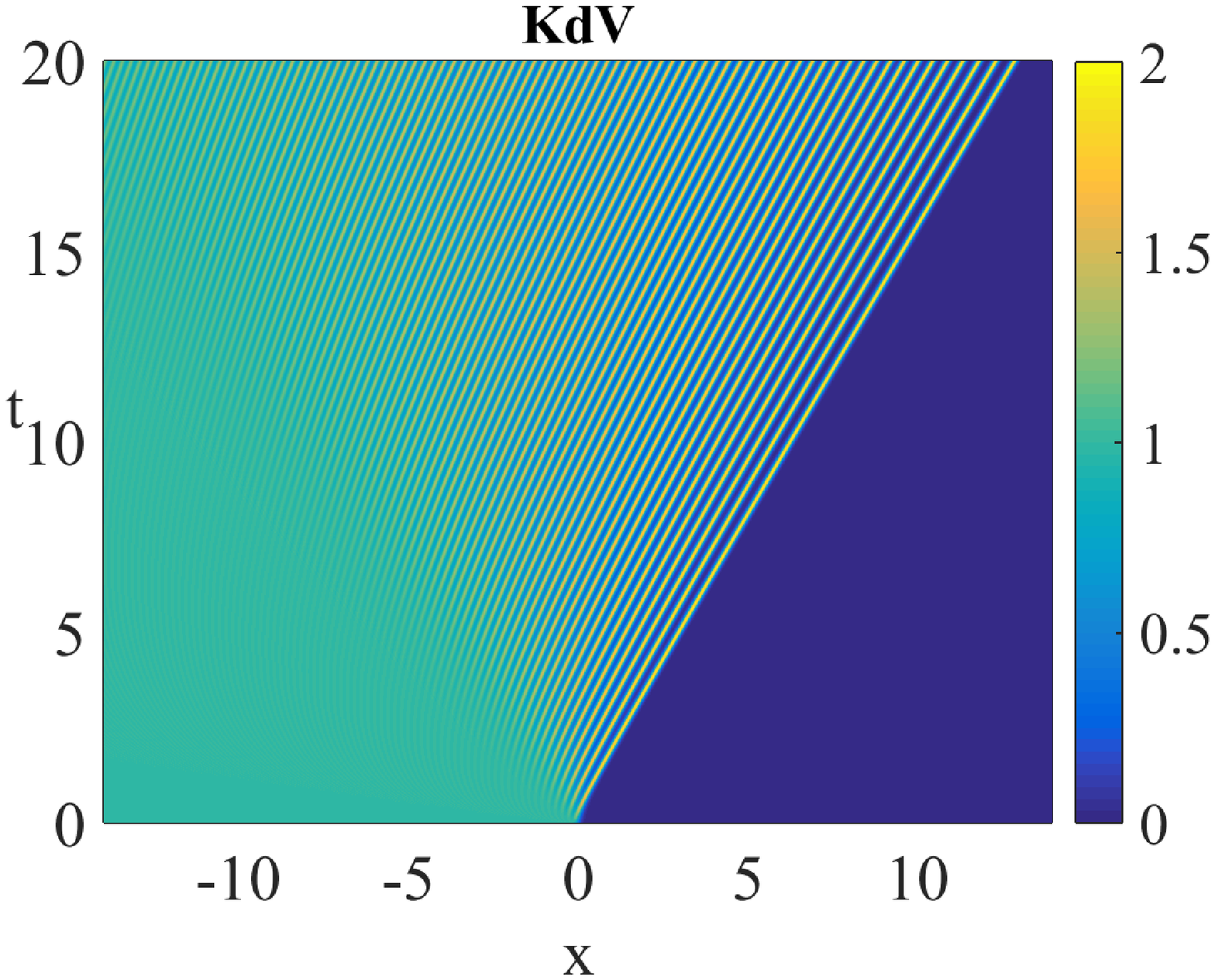}}
\subfigure[]{
\includegraphics[width=0.48\textwidth]{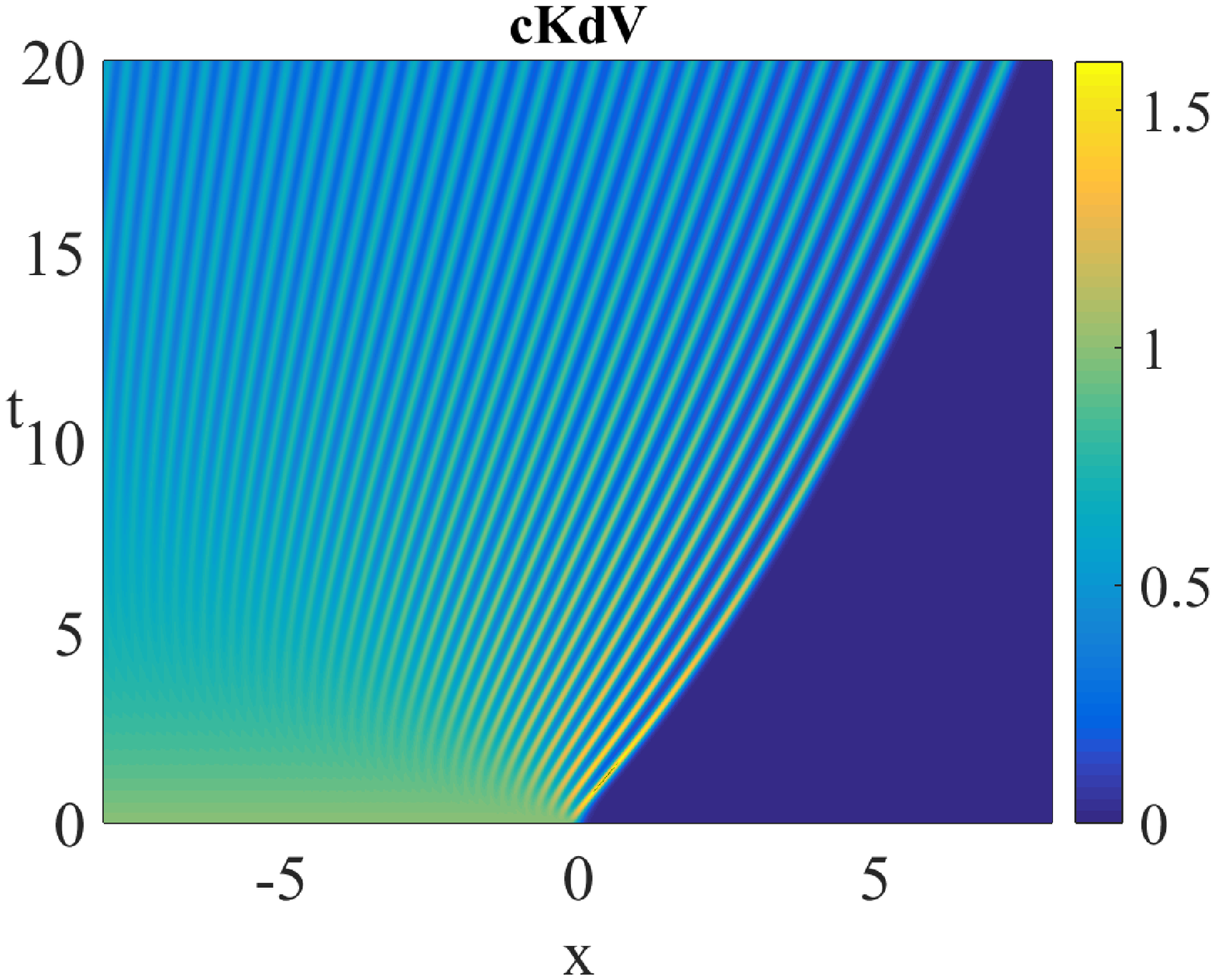}}
\caption{\small Space-time plot of the direct numerical solutions between $t=0$ and $t=20$ (a)for KdV eq., (b) for cKdV eq. Here, we take $t_0=10$ and $\epsilon^{2}=10^{-3}$.}
\label{fig5b}
\end{figure}

For both KdV and cKdV, the structure of the DSWs, its leading edge amplitudes and the wavelength of the oscillations predicted by the asymptotic solutions agree well with the numerical solutions. However in both cases, the position of the leading edges are slightly different, i.e. there is a small phase shift; this phase shift is larger towards the rear end of the DSW. Indeed one needs to proceed to higher order terms in the asymptotic expansion to achieve better results for the phase \cite{Abl70} and to smooth any shock-like discontinuities. This topic is outside the scope of this paper but we will return to it in a future communication. In our comparisons we have chosen the phase shift $\theta_0$ in the reconstructed leading order solution (\ref{soll}) to adjust the first maxima to agree with direct numerical simulations. This phase shift is not fixed due to the asymptotic nature of the initial value problem.

For the KdV equation, the amplitude of the trailing edge remains fixed at $1$, while the amplitude of the leading edge approaches $2$. The leading edge position is $\eta_+=4.946$ at $t=7.5$, so the average front speed of the DSW is calculated to be $V_+=0.659$. This speed agrees almost perfectly with the phase speed $V_+=2/3$ of the soliton solution of the KdV equation with amplitude $2$:
\begin{equation}
\label{solKdV}
f(\eta,t)=2\textrm{sech}^2\left[\frac{1}{\sqrt{6\epsilon^2}}\left(\eta-\eta_0-\frac{2}{3}t\right)\right].
\end{equation}

For the cKdV equation, the amplitude decays in time for the trailing edge. Accordingly the amplitude of the leading edge also decays in time, and so does the speed of the leading edge. Indeed in this case the leading edge position is $\eta_+=3.549$ at $t=7.5$, so the average front speed of the DSW is approximately $V_+=0.47$. This is smaller than the average front speed in the KdV case.

Finally we remark that the numerical solution  of the Whitham KdV equations in Fig.~\ref{fig3}a suggests that there is a self-similar solution of the reduced Whitham system (\ref{diag}) when $h_i=0$; indeed this is true and well-known. We note also that in general the Whitham system (\ref{diag}) associated with the cKdV equation has a similarity solution of the form $r_i(\eta,t)=r_i\left(\widetilde{\xi}\right), i=1,2,3$ where $\widetilde{\xi}=\eta/(2t+t_0)$; the similarity equations are given by
\begin{equation}
\label{sskdv}
\frac{\partial r_i}{\partial \widetilde{\xi}}\left(v_i\left(r_1,r_2,r_3\right)-2\widetilde{\xi}\right)+h_i\left(r_1,r_2,r_3\right)=0,\,\,\,\,\,\,\,\,\,i=1,2,3.
\end{equation}
However this similarity system is unlikely to be uniformly valid on the whole domain.

\section{Dispersive Shock Waves in the BO and cBO Equations }
In this section we will study DSWs associated with the BO and cBO equations. We will follow the  method described in the above section for the KdV and cKdV equations.

\subsection{Whitham modulation equations for BO/cBO -- conservation laws}

We begin with Eq.~(\ref{cBO}) and introduce the transformation of variables (\ref{vartransf})  into fast and slow coordinates. This yields the equation
 $$[(-\frac{\omega}{\epsilon} \frac{\partial}{\partial \theta} +  \frac{\partial}{\partial t}) + \epsilon \mathcal{H}  ( \frac{k}{\epsilon}  \frac{\partial}{\partial \theta} + \frac{\partial}{\partial \eta})^2]f+ f ( \frac{k}{\epsilon}  \frac{\partial}{\partial \theta}+ \frac{\partial}{\partial \eta}) f   + \frac{\lambda \tilde{c}}{1+2\lambda \tilde{c} t}f =0$$
or
\begin{equation}
\label{orderbo}
\begin{aligned}
&\frac{1}{\epsilon}\left[-\omega\frac{\partial f}{\partial \theta}+k f \frac{\partial f}{\partial \theta}+k^2\mathcal{H}\left(\frac{\partial^{2} f}{\partial \theta^{2}}\right)\right]\\
&+\left[\frac{\partial f}{\partial t}+f\frac{\partial f}{\partial \eta}+\mathcal{H}\left(k_{\eta}\frac{\partial f}{\partial \theta}+2k\frac{\partial^{2} f}{\partial \theta \partial \eta}\right)+\frac{\lambda \tilde{c}}{1+2\lambda \tilde{c}t}f\right]+\epsilon \mathcal{H}\left(\frac{\partial^2 f}{\partial \eta^{2}}\right)=0.
\end{aligned}
\end{equation}
We then expand $f=f_0+\epsilon f_1+...$ with the equations of the first two orders given at  $\mathcal{O}\left(\frac{1}{\epsilon}\right)$ by
\begin{equation}
\label{blo2}
-\omega f_{0,\theta}+k f_{0}f_{0,\theta}+k^{2}\mathcal{H}\left(f_{0,\theta\theta}\right)=0;
\end{equation}
and at $\mathcal{O}(1)$ by the linear equation
\begin{equation}
\label{bsop}
\mathcal{L}_{H}f_1\equiv-\omega f_{1,\theta}+k\left(u_{0}f_{1}\right)_{\theta}+k^{2}\mathcal{H}\left(f_{1,\theta\theta}\right)=G
\end{equation}
where
\begin{equation}
\label{gg2}
G\equiv-\left[f_{0,t}+f_{0}f_{0,\eta}+\mathcal{H}
\left(k_{\eta}f_{0,\theta}+2kf_{0,\theta\eta}\right)+\frac{f_0}{2t+t_0}\right].
\end{equation}
The solution of Eq.~(\ref{blo2}) is
\begin{equation}
\label{blos}
f_0\left(\theta,\eta,t\right)=\frac{4k^2}{\sqrt{A^2+4k^2}-A \cos(\theta-\theta_0)}+\beta
\end{equation}
where $\theta_0$ is constant.

Actually, Eq.~(\ref{blos}) is the periodic wave solution of the classical BO equation which was first obtained by Benjamin \cite{Ben67}. Here $A=\frac{1}{2}\left(f_{0,max}-f_{0,min}\right)$ is the amplitude of the wave (cf. \cite{Mat07}) and the phase velocity of the wave is given by
\begin{equation}
\label{pv}
V=\frac{1}{2}\sqrt{A^2+4k^2}+\beta.
\end{equation}
In Eq.~(\ref{blos}), $k$, $A$, $\beta$ and $V$
are functions of slow variables $\eta$ and $t$. In what follows we get the general modulation equations for the cBO equation in terms of the three variables $k$, $V$ and $\beta$; note from Eq.~(\ref{pv}) $A$ can be written in terms of these variables.

As with KdV and cKdV  the conservation of waves (\ref{comp}) is a necessary compatibility condition for the cBO equation as well.  This is the first conservation law. The other two conservation laws are obtained by eliminating the secular terms at the right-hand side of Eq.~(\ref{bsop}). In a similar manner to KdV/cKdV above, let $w$ denote solutions of the adjoint problem to $\mathcal{L}_{H}u=0$, i.e.,
\begin{equation}
\label{adj3}
\mathcal{L}_{H}^{A}w=0,\,\,\,\,\,\mathcal{L}_{H}^{A}= \omega \partial_{\theta}-k f_{0}\partial_{\theta}-k^{2}\mathcal{H}\left(\partial_{\theta\theta}\right)
\end{equation}
where we used the anti-symmetry of the Hilbert transform: $\langle\mathcal{H}u,v\rangle=\langle u,-\mathcal{H}v\rangle, \langle,\rangle$ being the standard inner product. In order to eliminate secular terms, we use the following relation that follows from (\ref{bsop})
\begin{equation}
\label{adj3}
\int_{0}^{2\pi}[w\mathcal{L}_{H}f_{1}-f_{1}\mathcal{L}_{H}^{A}w]d\theta=\int_{0}^{2\pi}wGd\theta.
\end{equation}
We put $w=1$ and $w=f_0$ (obtained from Eq.~(\ref{blo2})) into Eq.~(\ref{adj3}), enforce the periodicity of $f_0\left(\theta,\eta,t\right)$ in $\theta$ and obtain the secularity conditions respectively as
\begin{equation}
\label{bsecs}
\int_{0}^{2\pi}Gd\theta=0,\,\,\,\,\,\textrm{and}\,\,\,\,\int_{0}^{2\pi}f_{0}Gd\theta=0.
\end{equation}
Using following identity,
\begin{equation}
\label{bint1}
\int_{0}^{2\pi}\mathcal{H}\left(\frac{\partial f_0}{\partial \theta}\right)d\theta=0,
\end{equation}
we get from the first secularity condition in Eq.~(\ref{bsecs})
\begin{equation}
\label{biconl1}
\frac{\partial}{\partial t}\int_{0}^{2\pi}f_{0}d\theta+\frac{\partial}{\partial \eta}\left(\frac{1}{2}\int_{0}^{2\pi}f_{0}^{2}d\theta\right)
+\frac{1}{2t+t_0}\int_{0}^{2\pi}f_{0}d\theta=0
\end{equation}
and the second secularity condition in Eq.~(\ref{bsecs})
\begin{equation}
\label{biconl2}
\frac{\partial}{\partial t}\int_{0}^{2\pi}f_{0}^{2}d\theta+\frac{\partial}{\partial \eta}\left(\frac{2}{3}\int_{0}^{2\pi}f_{0}^{3}d\theta\right)+2\int_{0}^{2\pi}f_0\mathcal{H}
\left(k_{\eta}f_{0,\theta}+2kf_{0,\theta\eta}\right)d\theta+\frac{2}{2t+t_0}\int_{0}^{2\pi}f_{0}^{2}d\theta=0.
\end{equation}
From the properties of the Hilbert transform \cite{Ono75}, we have
\begin{equation}
\label{bident}
\begin{aligned}
&\int_{0}^{2\pi}f_{0}d\theta=2\pi\left(\beta+2k\right),\\
&\int_{0}^{2\pi}f_{0}^{2}d\theta=2\pi\left(\beta^{2}+4Vk\right),\\
&\int_{0}^{2\pi}f_{0}^{3}d\theta=2\pi\left[\beta^{3}+6\beta^{2}k+12k\beta\left(V-\beta\right)+k\left(3A^2+8k^2\right)\right],\\
&\int_{0}^{2\pi}f_{0}\mathcal{H}\left(k_{\eta}f_{0,\theta}+2kf_{0,\theta\eta}\right)d\theta=-\left[kA^2\right]_{\eta} .
\end{aligned}
\end{equation}
Substituting the definition (\ref{pv}) and identities (\ref{bident}) into Eqs.~(\ref{biconl1}) and (\ref{biconl2}) we can simplify them to find the following conservation laws
\begin{equation}
\label{bcclw2}
\beta_t+\beta\beta_{\eta}+\frac{2k+\beta}{2t+t_0}=0
\end{equation}
and
\begin{equation}
\label{bcclw3}
V_t+VV_{\eta}+kk_{\eta}+\frac{2V-\beta}{2t+t_0}=0.
\end{equation}
Equations~(\ref{comp}), (\ref{bcclw2}) and (\ref{bcclw3}) are the three conservation laws for the three variables $k$, $V$ and $\beta$.

\subsection{Whitham modulation equations for BO/cBO -- Riemann type variables}
It is convenient  to introduce  Riemann type variables $a,b,c$ \cite{Mat07}
\begin{equation}
\label{brv}
k=b-a,\,\,\,\,\,\,\,\,V=b+a,\,\,\,\,\,\,\beta=2c
\end{equation}
and write the leading order solution $f_0$ in terms of  $a,b,c$
\begin{equation}
\label{bsoll}
f_0\left(\theta,\eta,t\right)=\frac{2\left(b-a\right)^2}{\left(b+a-2c\right)-2\sqrt{\left(a-c\right)\left(b-c\right)}\cos(\theta-\theta_0)}+2c
\end{equation}
where the phase  $\theta$ is determined by Eq.~(\ref{iphase}) and $\theta_0$ is at this stage an arbitrary constant.

This transformation to Riemann type variables simplifies  the conservation laws.
We can write the quasilinear PDE system for Riemann variables $a$, $b$ and $c$ in the following form (recall $t_0=1/\tilde{c}$)
\begin{equation}
\label{bdiag}
\begin{aligned}
a_t+2aa_{\eta}+\frac{a+b-c}{2t+t_0}=0,\\
b_t+2bb_{\eta}+\frac{a+b-c}{2t+t_0}=0,\\
c_t+2cc_{\eta}+\frac{b+c-a}{2t+t_0}=0.
\end{aligned}
\end{equation}
The situation is similar to the KdV/cKdV case. In the absence of time dependent terms (formally $t_0 \rightarrow \infty$), Eq.~(\ref{bdiag}) reduces to a diagonal system.

We take initial values of $a,b,c$ to be
\begin{equation}
\label{bwic}
a(\eta,0) = \left\{
\begin{array}{lr}
 0, &  \eta \leq0;\\
 \frac{1}{2}, &   \eta >0,
 \end{array}
 \right.\,\,\,\,\,\,\,\,b(\eta,0)=\frac{1}{2},\,\,\,\,\,\,\,\,c(\eta,0)=0.
\end{equation}

\begin{figure}[ht]
\centering
\includegraphics[width=0.48\textwidth]{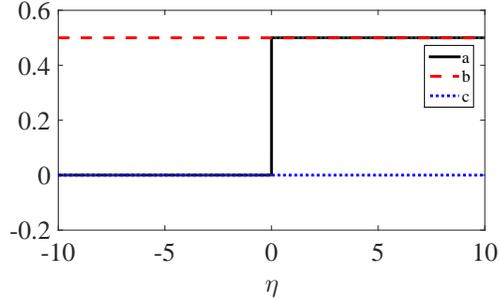}
\caption{\small Initial values (\ref{bwic}) for Riemann variables $a$, $b$ and $c$.}
\label{fig5A}
\end{figure}

Corresponding to the initial condition (\ref{bwic}) (see also Fig.~\ref{fig5A}), the Whitham system (\ref{bdiag})  for the classical BO equation (formally $t_0 \rightarrow \infty$)  admits an exact rarefaction wave solution  in terms of the self-similar variable $\xi=\eta/t$. It is $b=1/2, c=0$ and from
\begin{equation}
\label{r2sim}
(2 a-\xi) a'(\xi)=0
\end{equation}
we find $a=a(\xi)= \xi/2$ (cf. \cite{Mat98}); we also note that in \cite{Mat07}, stationary solutions of Whitham type systems for the BO-Burgers equation were investigated. Like the cKdV equation there is a self similar system of equations which can represent the solution of the cBO equation for a portion of the domain; this system is given by
\begin{equation}
\label{ssbo}
\begin{aligned}
2a_{\widetilde{\xi}}\left(a-\widetilde{\xi}\right)+a+b-c=0,\\
2b_{\widetilde{\xi}}\left(b-\widetilde{\xi}\right)+a+b-c=0,\\
2c_{\widetilde{\xi}}\left(c-\widetilde{\xi}\right)+b+c-a=0,
\end{aligned}
\end{equation}
where $\widetilde{\xi}=\eta/(2t+t_0)$.

As in the cKdV case, finding an analytical solution of  the general Whitham system (\ref{bdiag}) is more difficult because of its non-diagonal property.  Therefore we follow the  numerical approach developed for the cKdV case in order to understand the structure of DSWs in the cBO equation.

\subsection{Comparison between numerical simulations of Whitham modulation equations and direct numerical simulations of BO/cBO}

In order to numerically solve the  Whitham modulation equations in terms of Riemann type variables (\ref{bdiag})  we need to first obtain the boundary conditions. For the BO equation the boundary conditions are fixed in time and can be read from Eq.~(\ref{bwic}) or Fig.~\ref{fig5A}. In the case of the cBO equation (as opposed to the cKdV equation) we can find the boundary conditions for the Riemann variables analytically. Indeed, neglecting the derivatives with respect to the spatial variable $\eta$ in the Whitham system (\ref{bdiag}), we obtain a linear system whose solutions can be easily found. The exact solution for the boundary conditions on the left side  with  the initial conditions (\ref{bwic}) are given by
\begin{equation}
\label{bbcl}
\begin{aligned}
a_{-}&=\frac{R^2(t)}{2}+\frac{R(t)}{2}\left[1-R(t)\right]-\frac{1}{2},\\
b_{-}&=\frac{R^2(t)}{2}+\frac{R(t)}{2}\left[1-R(t)\right],\\
c_{-}&=\frac{1}{2}\left[R(t)-1\right]
\end{aligned}
\end{equation}
where $R(t)$ is given by (\ref{rr}). Similarly the boundary conditions on the right side corresponding to the initial conditions (\ref{bwic}) are found to be
\begin{equation}
\label{bbcr}
\begin{aligned}
a_{+}=b_{+}=\frac{R^2(t)}{2},\,\,\,\,\,\,\,\,\,c_{+}=0.
\end{aligned}
\end{equation}
As in the cKdV case, all Riemann variables for the cBO equation at the boundaries decay in time; see Fig.~\ref{fig12} for plots of these variables for $t \leq 30$.

\begin{figure}[ht]
\centering
\subfigure[]{
\includegraphics[width=0.48\textwidth]{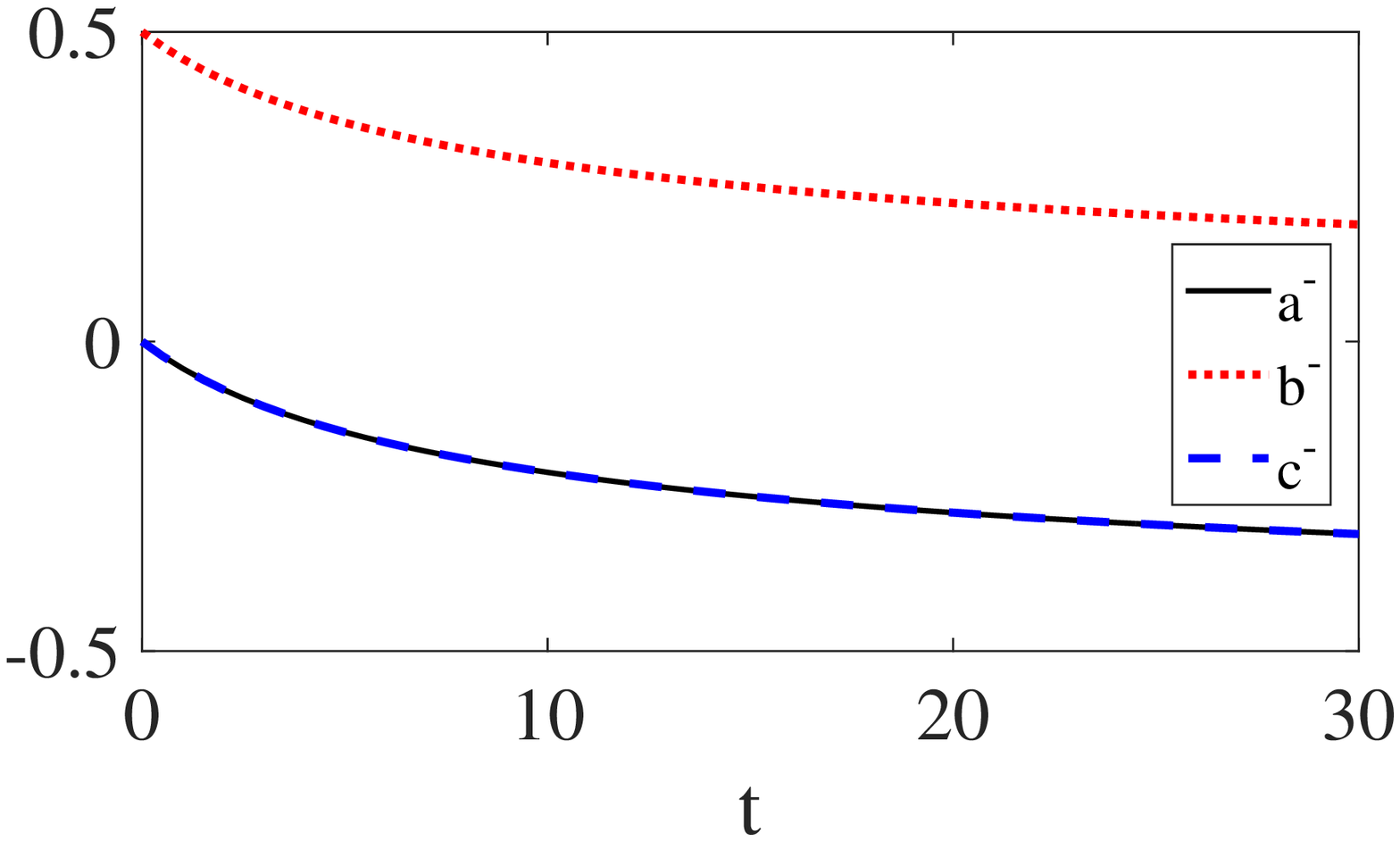}}
\subfigure[]{
\includegraphics[width=0.48\textwidth]{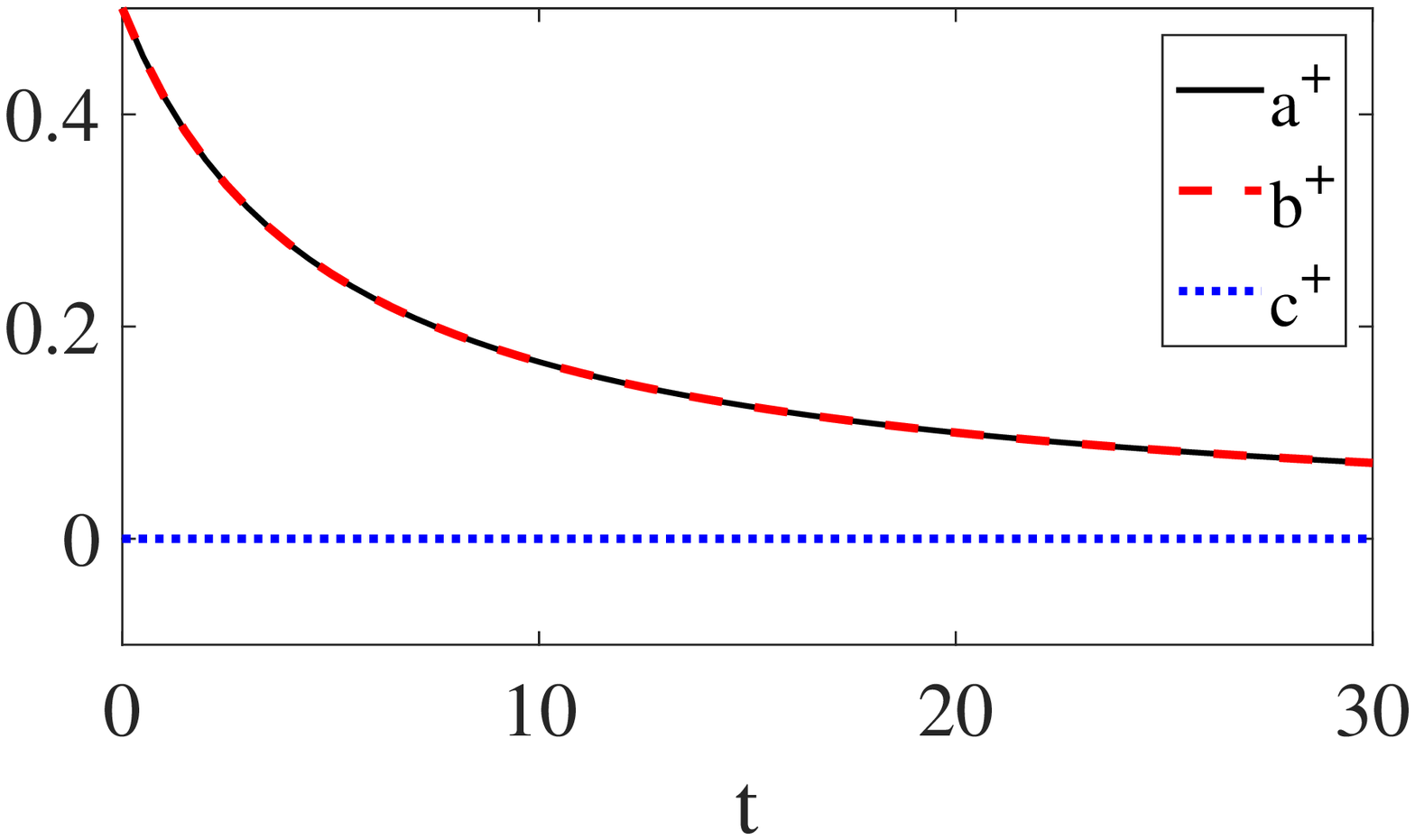}}
\caption{\small Evolution of Riemann variables for cBO case which are (a) given by (\ref{bbcl}) at the left boundary, (b) given by (\ref{bbcr}) at the right boundary.}
\label{fig12}
\end{figure}

In order to obtain the numerical solutions of the Whitham system (\ref{bdiag}), as with the KdV/cKdV case we again use Shampine's hyperbolic PDE solver with the same version of the Lax-Wendroff method. In the numerical computations, we use $N=2^{14}$ points for the spatial domain $[-30,30]$ with the time step being 0.9 times the spatial step. The results computed with the initial condition (\ref{bwic}) and the boundary conditions given above are shown in Fig.~\ref{fig8A} at time $t=7.5$. We note that there appear to be derivative discontinuities  at the leading and trailing edges of the DSW. These may be smoothed by keeping higher order terms, but doing so is outside the scope of this paper.

\begin{figure}[ht]
\centering
\subfigure[]{
\includegraphics[width=0.48\textwidth]{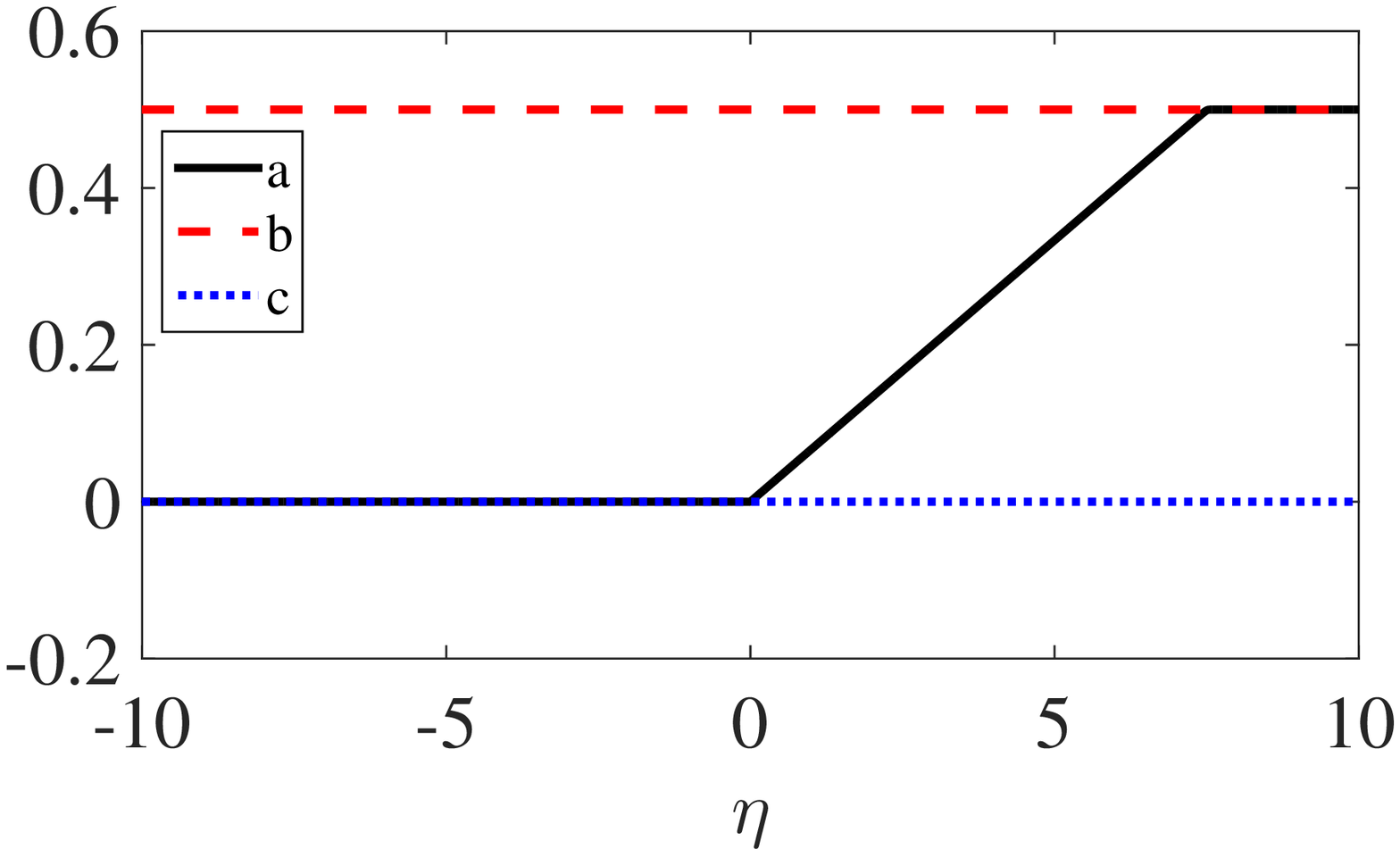}}
\subfigure[]{
\includegraphics[width=0.48\textwidth]{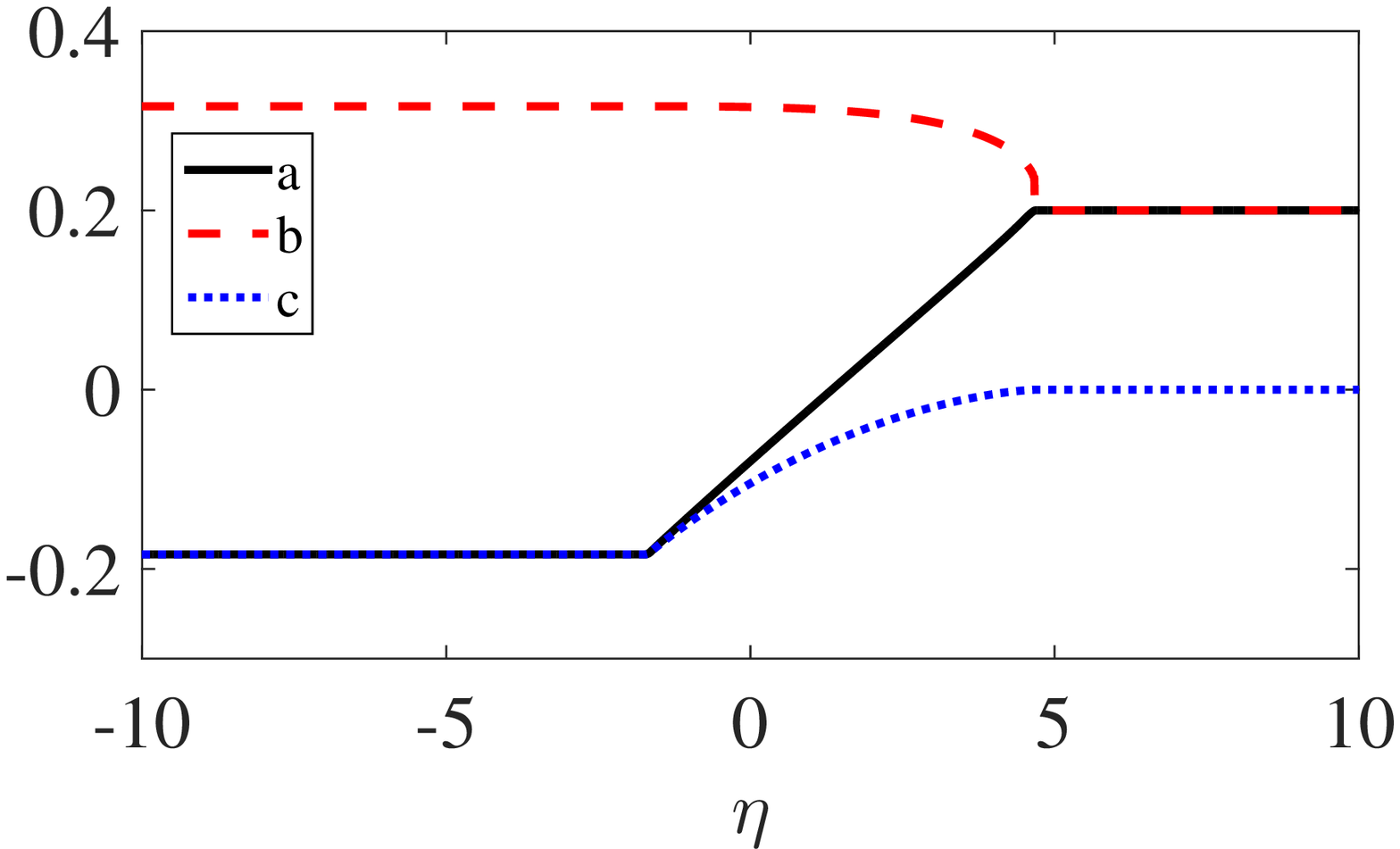}}
\caption{\small Riemann variables at t=7.5 which are found by numerical solutions of reduced Whitham system for BO eq. and exact Whitham system (\ref{diag}) for cBO eqs. (a) for BO eq., (b) for cBO eq. Here, we take $t_0=10$.}
\label{fig8A}
\end{figure}

We can reconstruct asymptotic solutions of the DSWs for both BO and cBO at any time $t$ from the numerical solutions of the Riemann variables using Eqs.~(\ref{brv}) and (\ref{bsoll}) with $\theta$ given by Eq.~(\ref{iphase}). These reconstructed solutions are compared with direct numerical simulations in Fig.~\ref{fig6}.

\begin{figure}[ht]
\centering
\subfigure[]{
\includegraphics[width=0.48\textwidth]{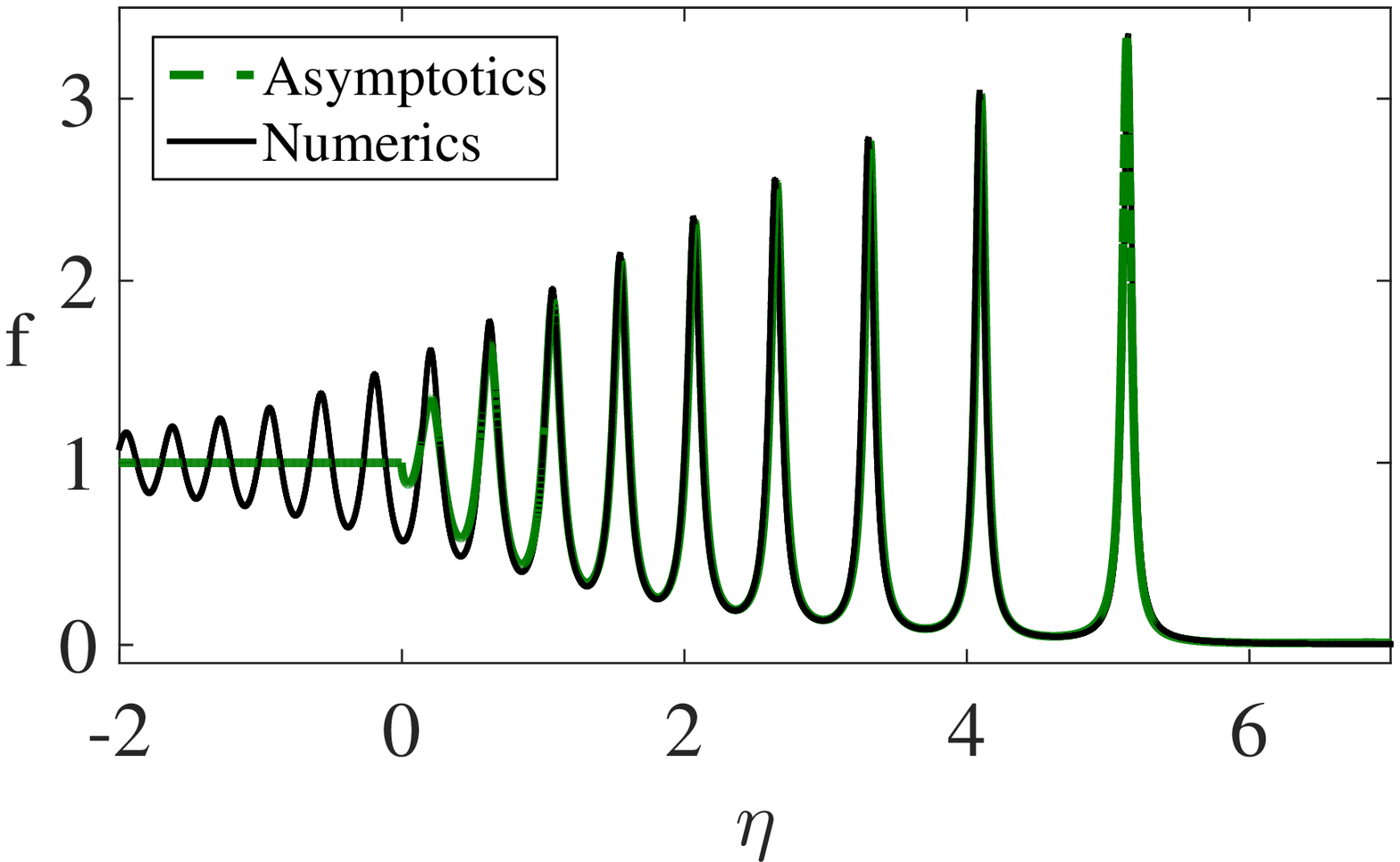}}
\subfigure[]{
\includegraphics[width=0.48\textwidth]{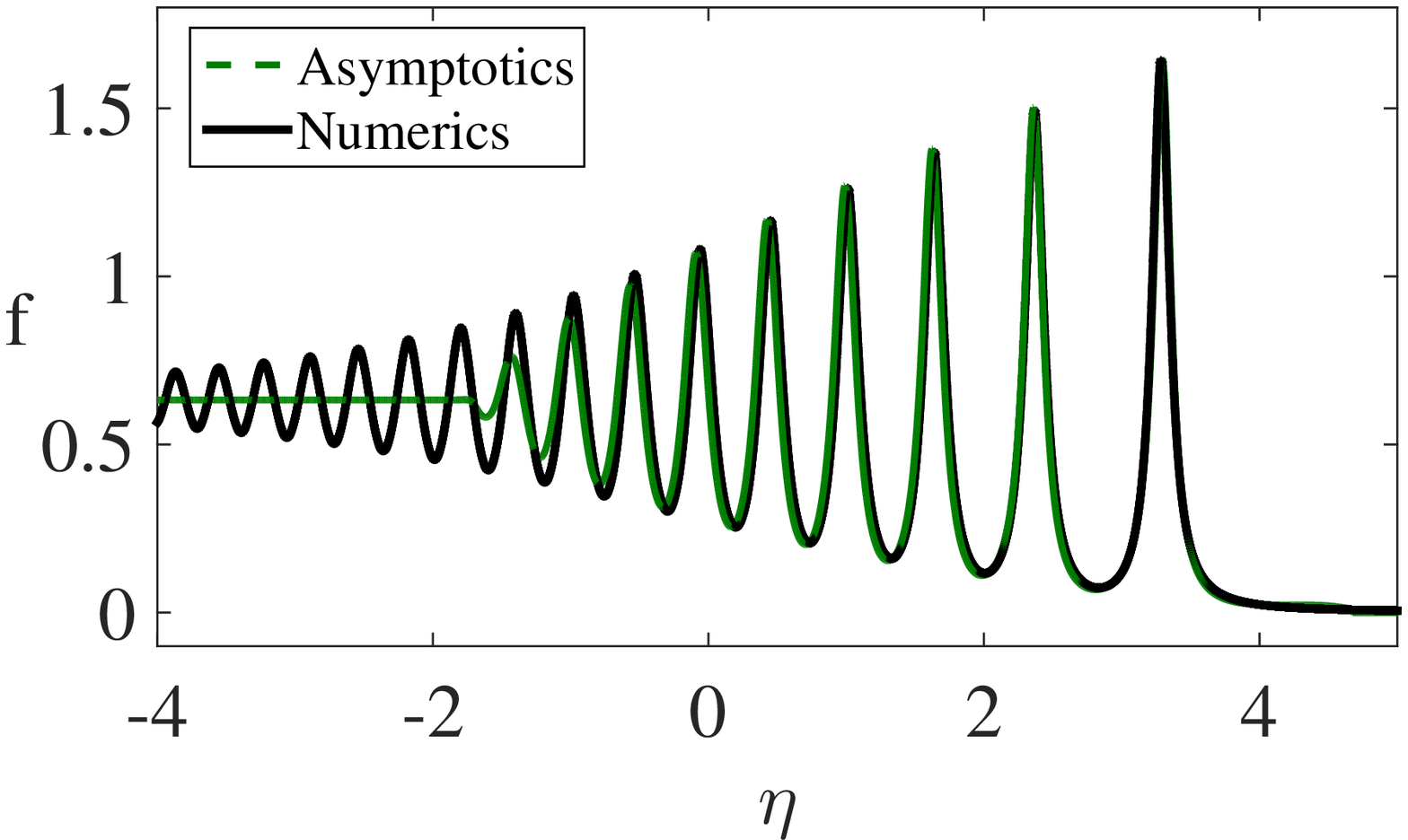}}
\caption{\small Numerical and asymptotic solutions of BO and cBO eqs. at t=7.5 with the initial data (\ref{ric}). (a)for BO eq., (b) for cBO eq. Here, we take $t_0=10$ and $\epsilon=10^{-3/2}$.}
\label{fig6}
\end{figure}

Next we solve the BO and cBO  equations (i.e.~Eq.~(\ref{cBO}) with $\tilde{c}=0$ and $\tilde{c} \neq 0$ respectively) numerically with the initial condition (\ref{ric}) regularized as in Eq.~(\ref{sic}). The numerical method used is completely analogous to KdV/cKdV, except that $\epsilon^2 \partial_{\eta\eta\eta}$ is replaced by $\epsilon \mathcal{H} \partial_{\eta\eta}$. In this computation, we use $N=2^{14}$ spatial Fourier modes with the domain size $L=30$, and choose the time step to be $10^{-4}$. The regularization parameter in the initial condition (\ref{sic}) is chosen to be $\tilde{K}=10$, and the parameters in the cBO equation (\ref{cBO}) are taken to be $t_0=10$ and $\epsilon=10^{-3/2}$.  Direct numerical simulations of BO/cBO and solutions of the associated Whitham equations are compared in Fig.~\ref{fig6} at $t=7.5$. In the reconstructions from the Whitham equations, we fix the arbitrary constant phase $\theta_0$ in Eq.~(\ref{bsoll}) by adjusting the maxima of the Whitham reconstruction to agree with those from direct numerical simulations.  We provide another perspective by including space-time plots of the direct numerical solutions of BO and cBO eqs.-- see Fig.\ref{fig6b}.

From Fig. \ref{fig6} it is  clear that the results of direct numerical simulations and those of the Whitham equations are overall in very good agreement. While the humps in the leading edge of the DSW are captured nearly perfectly,  we note that at the trailing edge there are discrepancies. This is due to the fact that as time increases the DSW humps move to the right and spread apart (unlike KdV). This phenomena is also observed in Fig.\ref{fig6b} for both BO and cBO eqs. For large time the trailing edge of the DSW becomes small and we expect higher order terms in the Whitham modulation equations will need to play a role. As with determining the arbitrary phase $\theta_0$, the higher order analysis is outside the scope of this paper.

\begin{figure}[ht]
\centering
\subfigure[]{
\includegraphics[width=0.48\textwidth]{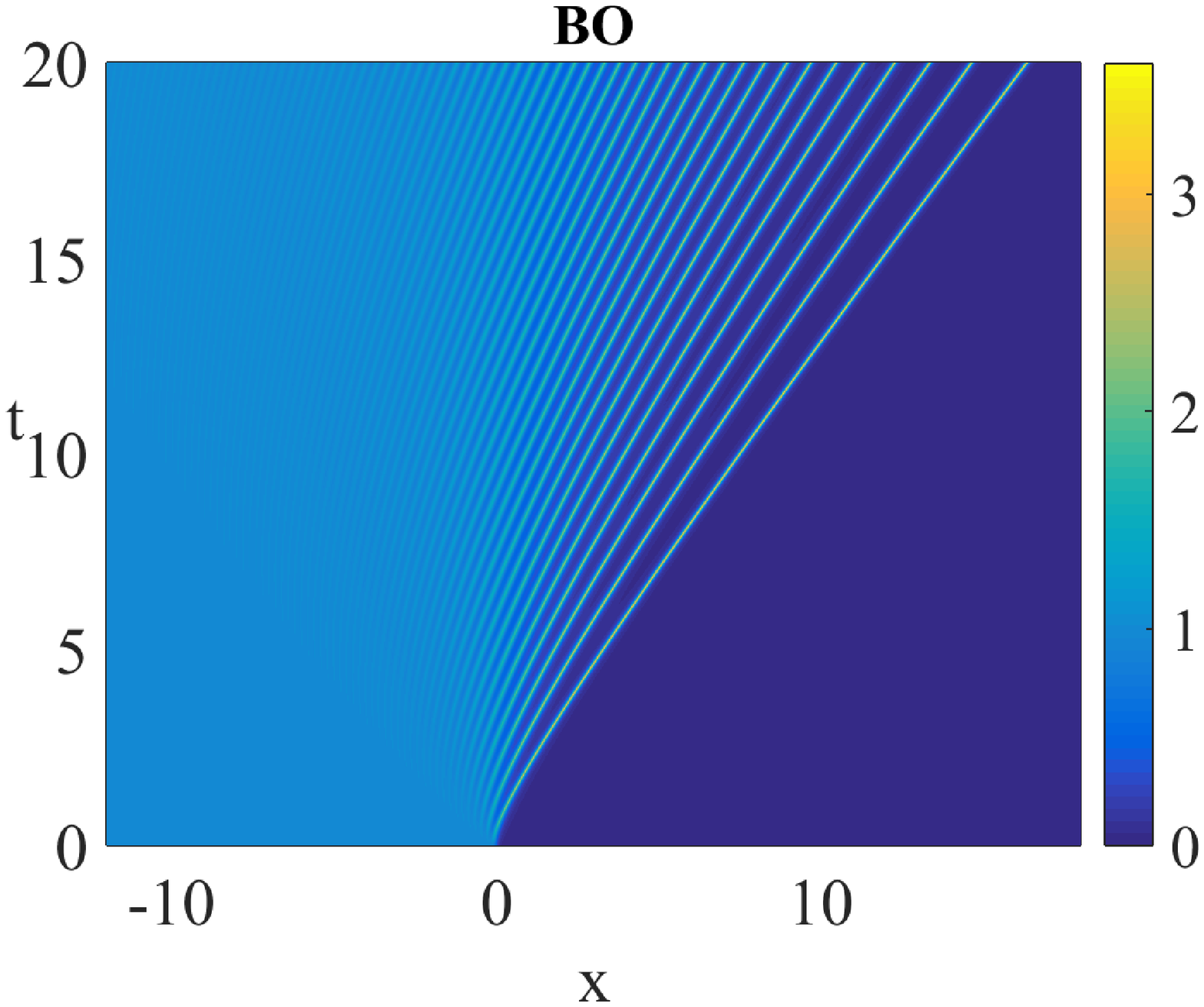}}
\subfigure[]{
\includegraphics[width=0.48\textwidth]{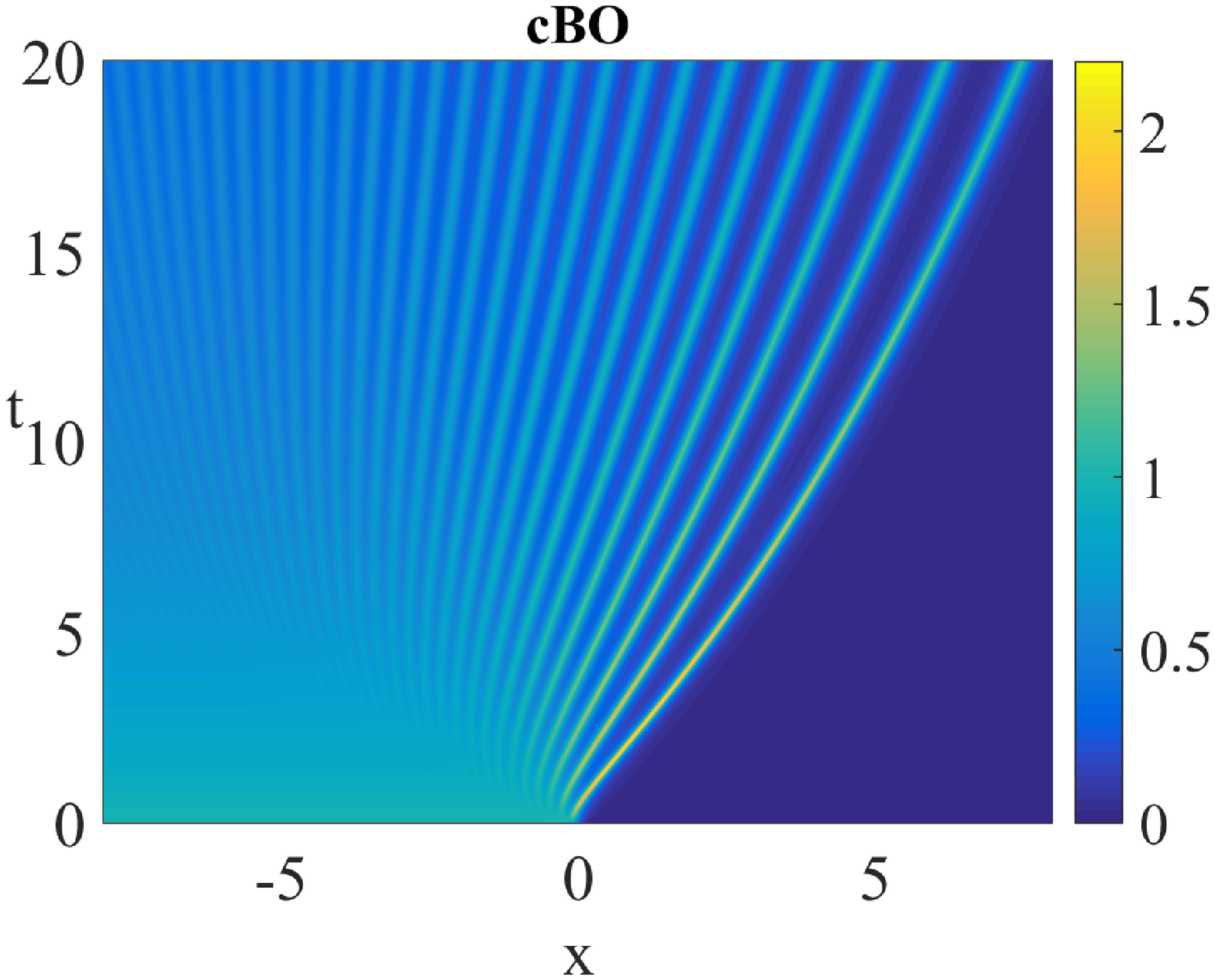}}
\caption{\small Space-time plot of the direct numerical solutions between $t=0$ and $t=20$ (a)for BO eq., (b) for cBO eq. Here, we take $t_0=10$ and $\epsilon=10^{-3/2}$.}
\label{fig6b}
\end{figure}

For the BO equation, we observe that the amplitudes of the leading hump increase slowly in time; they are calculated to be $3.431$, $3.586$ and $3.648$ for $t=10$, $20$ and $30$, respectively. This suggests that as $t \rightarrow \infty$ the amplitude may asymptote to 4.
The average speed of the leading hump of the DSW is approximately $V_{avg}=0.831$ at $t=7.5$. This speed is close to the phase speed of the algebraic solitary wave solution of the BO equation \cite{Ono75} with an amplitude $4V_{avg}=3.34$; this solitary wave is approximately represented by
\begin{equation}
\label{solBO}
f(\eta,t)=\frac{4V}{1+\left[\frac{V\left(\eta-Vt\right)}{\epsilon}\right]^2}.
\end{equation}

For the cBO equation, the average speed of the leading hump of the DSW at $t=7.5$ is approximately $V_{avg}=0.438$. This speed is considerably smaller  than in the BO case because the amplitude of the leading edge decreases in time. The  trailing edge of the DSW looks similar to that of the BO equation but its amplitude also decreases in time (see Fig.~\ref{fig6}). These features are similar to those observed for DSWs in the cKdV equation.

\section{Comparison with direct numerical simulations of KP/2DBO}
In this section we will solve the KP and 2DBO equations numerically using a modified version of Trefethen's code (Program 27 in \cite{Tre00})  and compare with direct numerical simulations of the corresponding (1+1) cylindrical equations.
First Eq.~(\ref{geq}) is written in 2D Fourier space as
\begin{equation}
\label{fgeq}
\widehat{u}_{t}+\widehat{\mathcal{L}}\widehat{u}+\frac{i}{2}k_x\widehat{u^2}=0.
\end{equation}
Here $\widehat{u}$ and $\widehat{u^2}$ are Fourier transforms of $u(x,y,t)$ and $u^2(x,y,t)$, respectively and the form of the operator $\widehat{\mathcal{L}}$ for the KP and 2DBO equations are
\begin{equation}
\label{opkp}
\widehat{\mathcal{L}}=\frac{i \lambda k_{y}^{2}}{k_x+\widehat{\lambda}}-i\epsilon^{2}k_{x}^{3}
\end{equation}
and
\begin{equation}
\label{opbo}
\widehat{\mathcal{L}}=\frac{i \lambda k_{y}^{2}}{k_x+\widehat{\lambda}}-i\epsilon \textrm{sgn}(k_x)k_{x}^{2}.
\end{equation}
We note that a regularization parameter $\widehat{\lambda}$ is added to the denominator $\widehat{\mathcal{L}}$  to prevent the singularity in (\ref{fgeq}) near $k_x=0$. Then by using integrating factor $e^{t\widehat{\mathcal{L}}}$, Eq.~(\ref{fgeq}) is written in the following equivalent form
\begin{equation}
\label{fgeq2}
\left(e^{t\widehat{\mathcal{L}}}\widehat{u}\right)_{t}+\frac{i}{2}k_xe^{t\widehat{\mathcal{L}}}\widehat{u^2}=0.
\end{equation}
Finally, a fourth order Runge-Kutta method is used for time integration of Eq.~(\ref{fgeq2}). For Fourier spectral methods, the initial condition must be periodic and smooth. However, the parabolic front initial condition (\ref{ic}) does not satisfy these conditions. Therefore we use the following regularized initial condition instead of Eq.~(\ref{ic}) in the numerical computations
\begin{equation}
\label{nic}
u_{I}(x,y,0)=\frac{1}{2}\left[\mu\, \textrm{tanh}\left(K\left(x+\frac{P(y,0)}{2}\right)\right)-\mu\, \textrm{tanh}\left(K\left(x+l_0+\frac{\tilde{P}(y,0)}{2}\right)\right)\right]\exp\left(-m^p\left|\frac{2y}{L_y}\right|^p\right).
\end{equation}
Here the 2D computational domain is $\left[-L_x,L_x\right]\times\left[-L_y,L_y\right]$.
This initial condition decays for large $x$; consequently we have effectively imposed  periodic boundary conditions. We choose the location of the backward front  to be far from the forward front and  the curvature of the backward front $\tilde{P}(y,0)$ to be much smaller than the curvature of the forward front $P(y,0)$; this minimizes the effect of the backward front. The parameter $K$ regularizes the  fronts (forward and backward), while the parameters  $m$ and $p$  smooth the initial condition in the $y$-direction (see Fig.~\ref{fig8a}).

\begin{figure}[ht]
\centering
\includegraphics[width=0.48\textwidth]{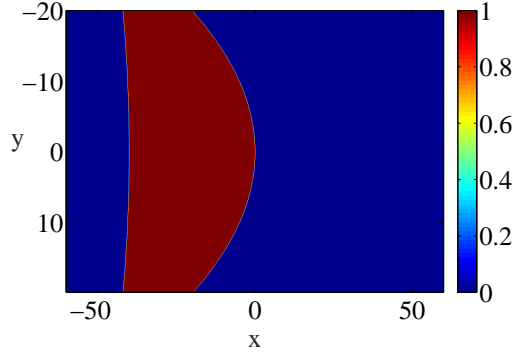}
\caption{\small Contour plot of the numerical initial condition (\ref{nic}) for $\mu=-1, K=10, P(y,0)=0.1y^2, l_0=40, \tilde{P}(y,0)=0.01y^2, m=0.05, p=3, L_x=60$ and $L_y=20$.}
\label{fig8a}
\end{figure}

In all numerical simulations we use large domain sizes $L_x$ and $L_y$, and an initially parabolic front $P(y,0)=\tilde{c} y^2$ with $\tilde{c}=0.1$. We choose the regularization parameters  $K$, $l_0$, $m$ and $p$ such that the relevant dynamics are locally equivalent between the exact initial condition (\ref{ic}) and the numerical initial condition (\ref{nic}).

The regularization parameter $\widehat{\lambda}$ that prevents the singularity at $k_x=0$ in Eq.~(\ref{fgeq}) is chosen to be the complex number $\widehat{\lambda}=i \widehat{\lambda_0}$, where $\widehat{\lambda_0}=2.2204\times10^{-16}$ is the smallest floating number in computations which MATLAB allows \cite{Kle07}. Meanwhile, in Fourier space, we have $\widehat{u}_{I}\left(0,k_y\right)\neq0$ for all  $k_{y} \in \mathbb{R}$ for the numerical initial condition (\ref{nic}) (this is also true for the exact initial condition (\ref{ic})). Even though the term $e^{t\widehat{\mathcal{L}}}$ near $k_x=0$ is a very small  number, this nevertheless has an effect on the zero background of the solution which is the natural background of the solution: the magnitude of the background changes with time $t$. Therefore, to compare with the numerical solutions of $1+1$ dimensional cylindrical equations, we readjust the numerical solutions of the $2+1$ dimensional equations to an appropriate background when we compare solutions.

For the KP equation, we choose the spatial resolution to be $2^{14}\times2^{10}$, the domain size to be $L_x=60$ and $L_y=20$, and the time step to be $10^{-4}$. We use the parameters $K=10, l_0=40, m=0.05, p=3, \tilde{P}(y,0)=0.01y^{2}$ and $\epsilon^{2}=10^{-3}$. The numerical solutions of the cKdV equation with the ETDRK4 method and the KP-II equation with Trefethen's code at $y=0$ and $y=\pm1.25$ are compared in Fig.~\ref{fig8} for $t=7.5$. To provide another perspective, we give a contour plot of the numerical solution of the KP-II equation --see Fig.\ref{fig8b}; also provided is an animation  of the DSW propagation in KP-II between $t=0$ and $t=8$ (see \cite{Mov1}).

\begin{figure}[ht]
\centering
\subfigure[]{
\includegraphics[width=0.45\textwidth]{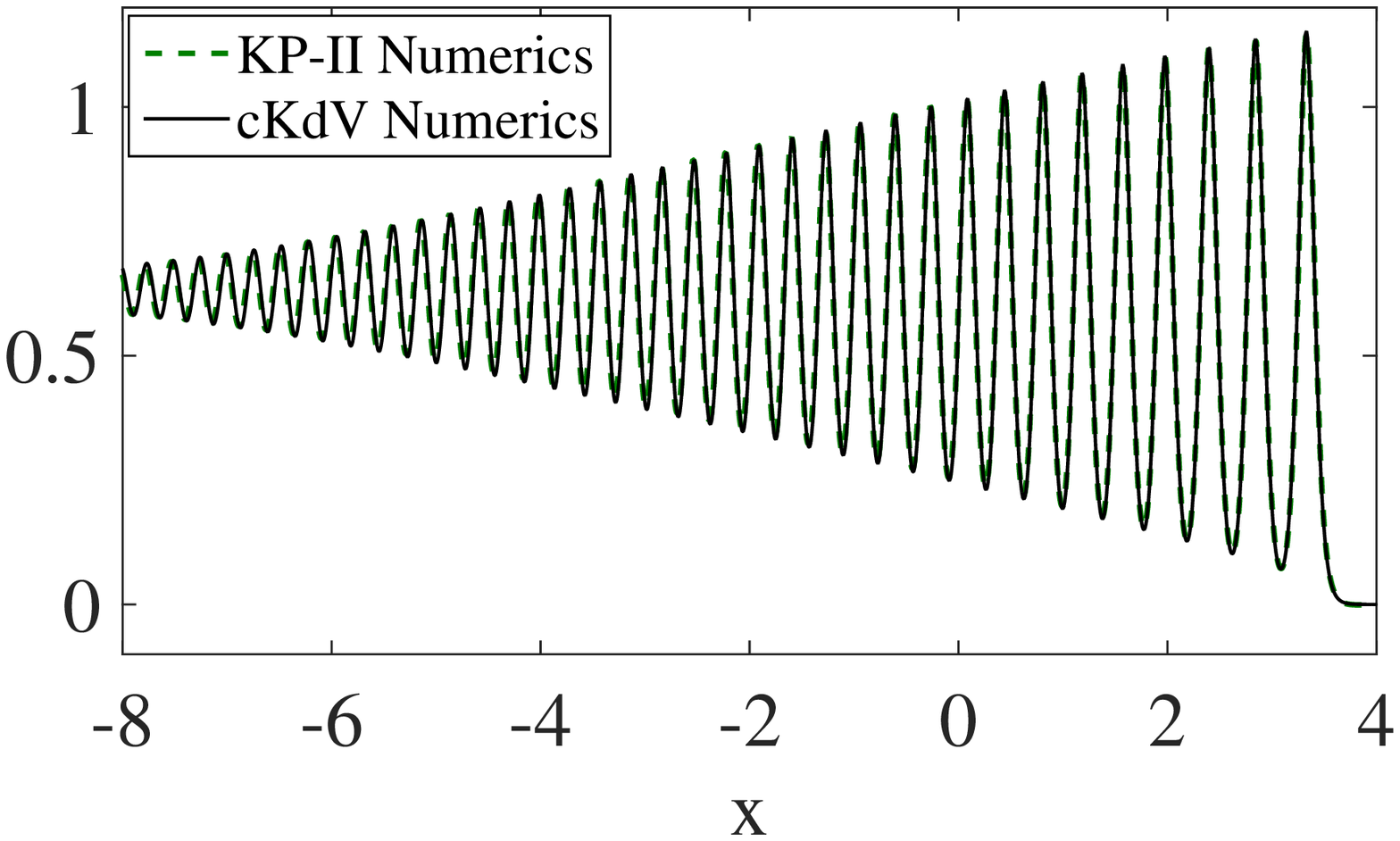}}
\subfigure[]{
\includegraphics[width=0.45\textwidth]{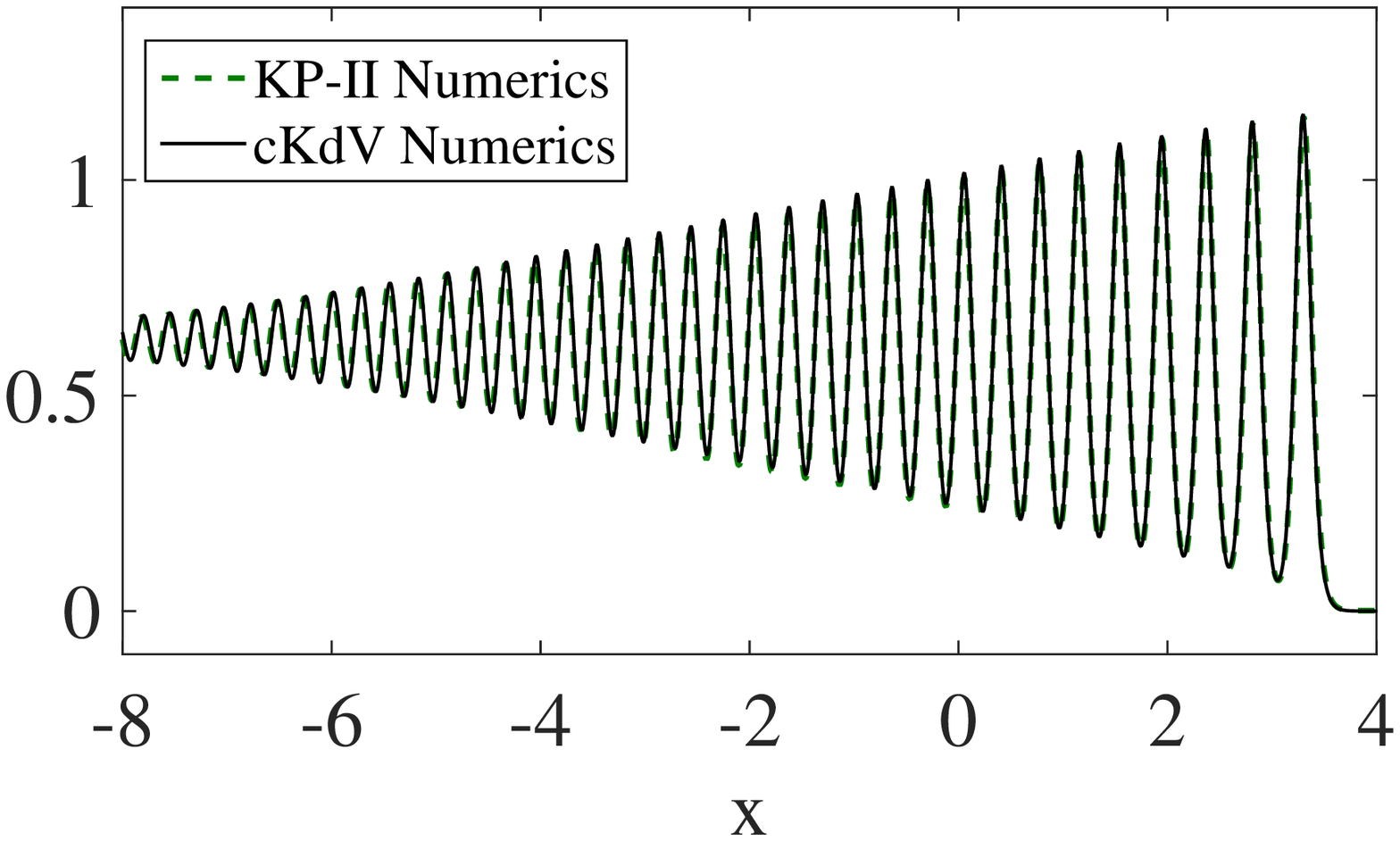}}
\caption{\small . Comparison of numerical solution of cKdV equation and numerical solution of KPII equation for t=7.5 (a) at $y=0$, (b) at $y=\pm1.25$. Here, we take $t_0=10$ and $\epsilon^{2}=10^{-3}$.}
\label{fig8}
\end{figure}

The solution of the corresponding $1+1$ dimensional cylindrical equation coincides with the solution of the 2+1 dimensional equation only at $y=0$. For the comparison of results at cross sections different from $y=0$, the solution of the 1+1 dimensional cylindrical equation must be shifted in the horizontal direction by a value which can be determined by the solution of the FS equation (\ref{center}) for the desired $y$-cross section and time $t$. Specifically we recall $\eta=x+P(y,t)/2$, where $P(y,t)=\frac{\tilde{c}y^2}{1+2\tilde{c} \lambda t}$ with $\lambda=1, \tilde{c}=0.1$.
Therefore, for the comparison of solutions of cKdV and KP-II equations, shifting is performed by $x=\eta-0.03125$ for $y=\pm1.25$ and $t=7.5$.

DSWs in KP-II equation are observed in both Fig.\ref{fig8b} and the animation given in \cite{Mov1}.

\begin{figure}[ht]
\centering
\includegraphics[width=0.48\textwidth]{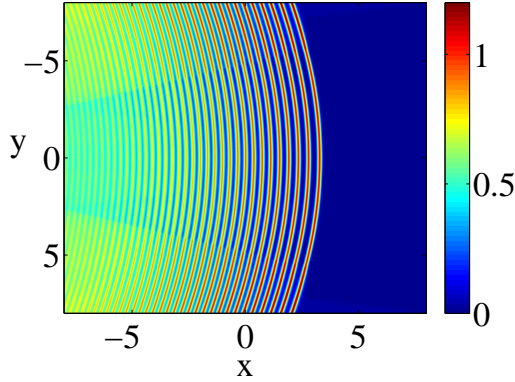}
\caption{\small Contour plot of the numerical solution of the KP-II equation at $t=7.5$ with computation parameters for the KP-II eq. given in the text.}
\label{fig8b}
\end{figure}

For the 2DBO equation, we choose the spatial resolution to be $2^{14}\times2^{10}$, the domain size to be $L_x=50$ and $L_y=20$, and the time step to be $10^{-4}$.} We use the parameters $K=10$, $l_0=30, m=0.05, p=3$, $\tilde{c}=0.1$  and $\epsilon=10^{-3/2}$.

\begin{figure}[ht]
\centering
\subfigure[]{
\includegraphics[width=0.48\textwidth]{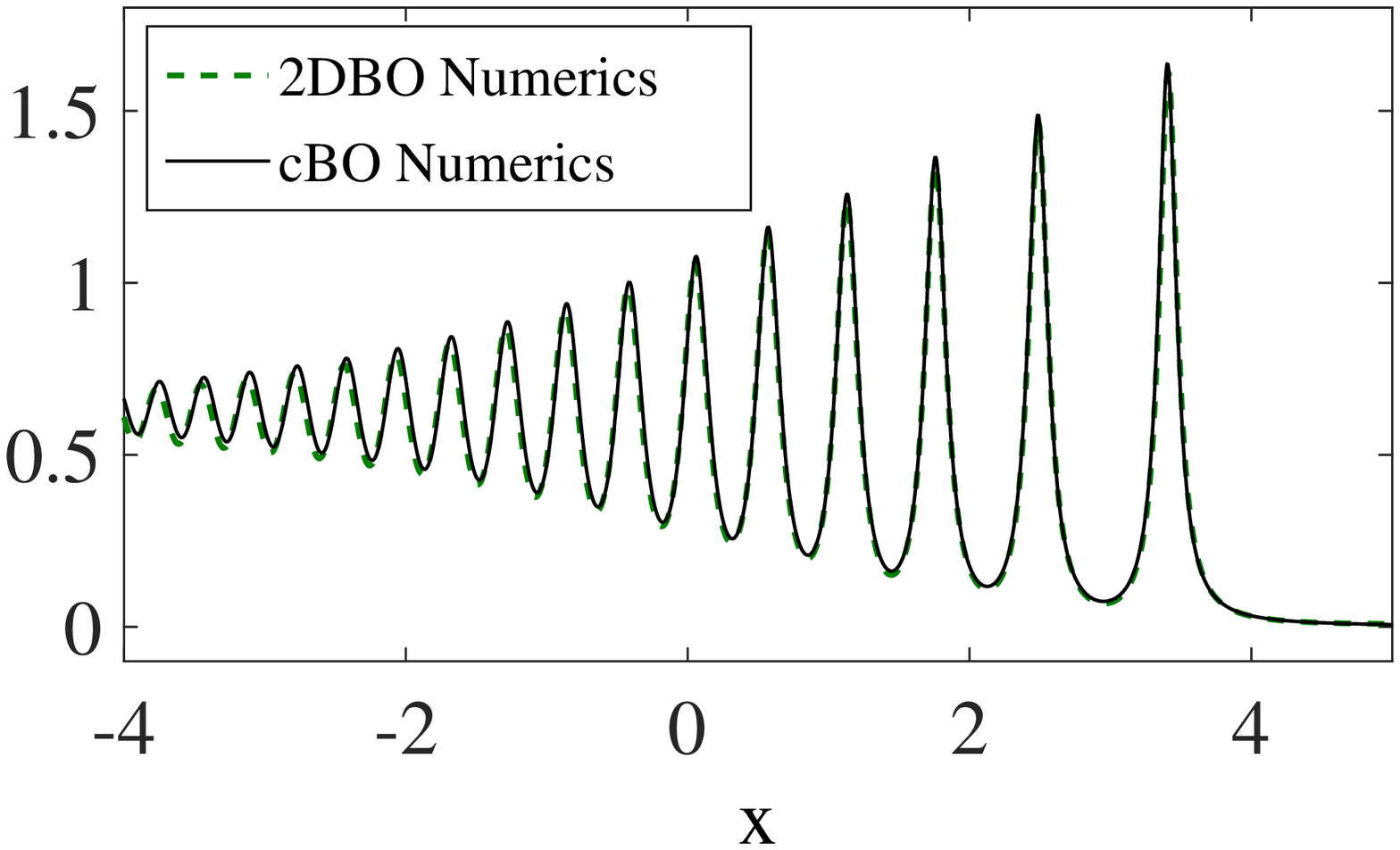}}
\subfigure[]{
\includegraphics[width=0.48\textwidth]{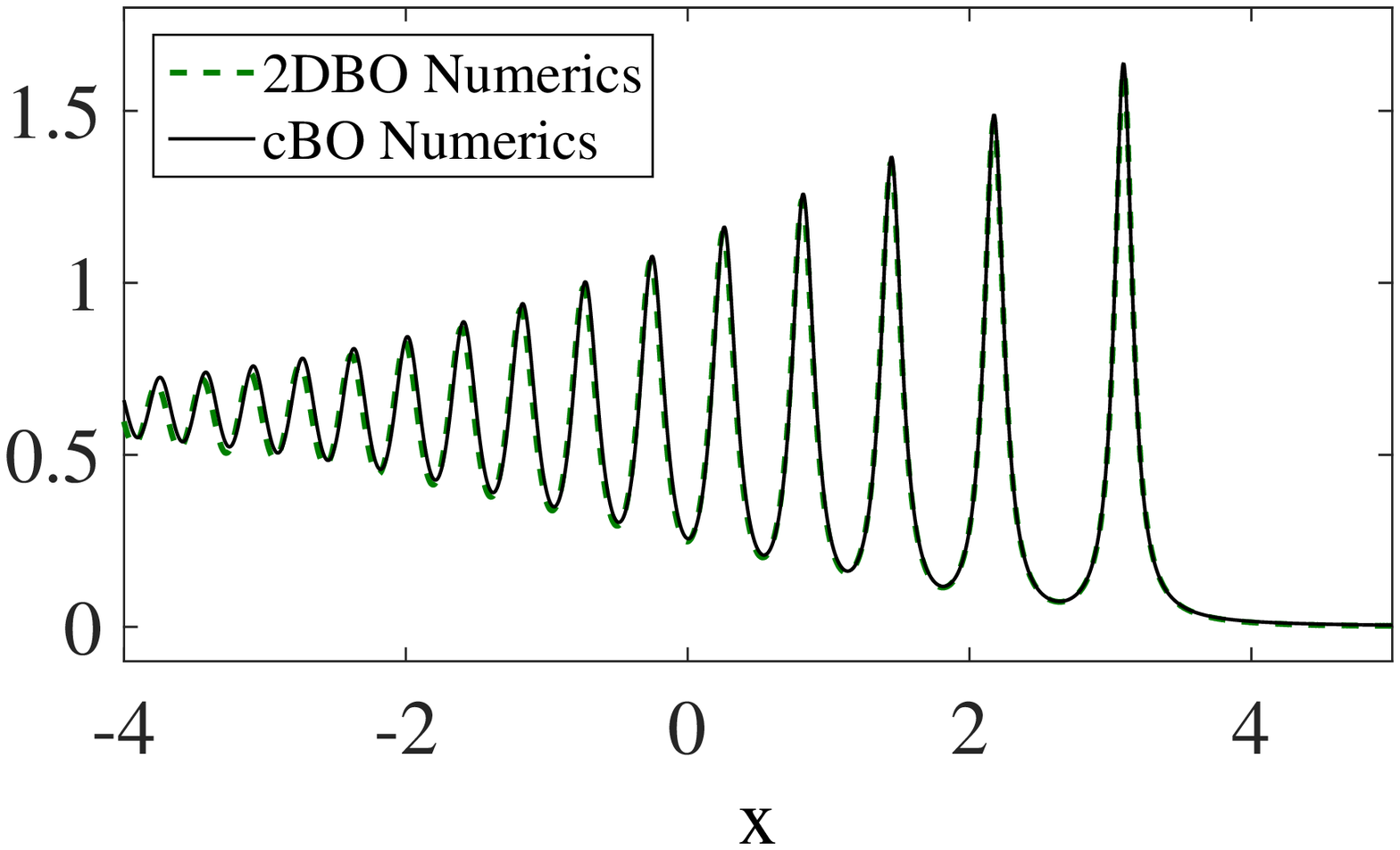}}
\caption{\small . Comparison of numerical solution of cBO equation and numerical solution of 2DBO equation for t=7.5 (a) at $y=0$, (b) at $y=\pm1.25$. Here, we take $t_0=10$ and $\epsilon=10^{-3/2}$.}
\label{fig9}
\end{figure}

\begin{figure}[ht]
\centering
\includegraphics[width=0.48\textwidth]{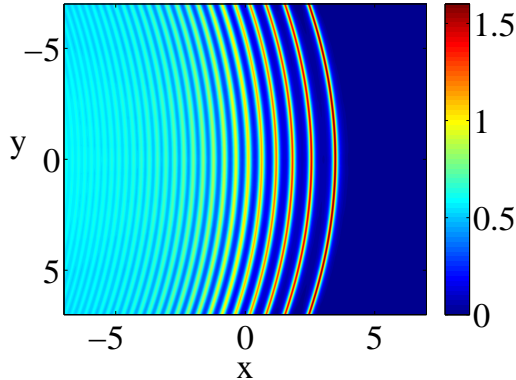}
\caption{\small Contour plot of the numerical solution of the 2DBO equation at $t=7.5$ with computation parameters for the 2DBO eq. given in the text.}
\label{fig9b}
\end{figure}

The numerical solution of the cBO equation with ETDRK4 method and the 2DBO equation with Trefethen's code at $y=0$ and $y=\pm1.25$ are compared in Fig.~\ref{fig9} for $t=7.5$. Similarly in KP-II, contour plot of the numerical solution of the 2DBO equation at $t=7.5$ given in Fig.\ref{fig9b} and propagation of DSWs in 2DBO between $t=0$ and $t=8$ obtained by the numerical solution is achieved (see \cite{Mov2} for the animation).

Here shifting is also performed by $x=\eta-0.03125$ for $y=\pm1.25$ and $t=7.5$; this is because the evolution of the front shape as given by the FS equation (\ref{center}) are identical between KP and 2DBO. DSWs in 2DBO equation are observed in both Fig.\ref{fig9b} and the animation given in \cite{Mov2}. The spreading behavior of DSW humps of BO type equations that we mentioned section 4.3 is also observed for the time evolution of 2DBO eq. in the animation \cite{Mov2}.

In both comparisons, there is excellent agreement between solutions of the 2+1 dimensional equations and the 1+1 dimensional cylindrical equations at the chosen $y$-cross sections and time. This supports Whitham's asymptotic method and the 1+1 dimensional cylindrical equations  as accurate reductions of the $2+1$ dimensional equations.Indeed in terms of computational time, the numerical solutions of 2+1 dimensional equations, 1+1 dimensional equations and Whitham systems are on the order of days, hours and minutes, respectively. Hence, in addition to enhanced understanding of the underlying DSWs, the asymptotic method has enormous advantages in terms of computational time.

We note that the level of agreement between 2+1 and 1+1 numerical solutions at subsequent  times depends on the $y$-domain size $L_y$ in the $2+1$ dimensional equations. If the $y$-boundaries are relatively far away from $y=0$, then initially the DSW does not hit the $y$-boundaries. However after a finite amount of time, the DSW reaches the $y$-boundaries and gets reflected. To delay this reflection effect, $L_y$ and correspondingly the number of grid points in $y$ need to increase.

Another aspect of the numerical solutions of 2+1 dimensional equations to note is the evolution in the $x$-direction. Since we use the regularized initial condition (\ref{nic}), as time evolves oscillations in $x$ exist on the left tail of the solutions. After some time, these oscillations affect the localized DSW behavior of the 2+1 dimensional equation  and thus agreement with the solution of the 1+1 dimensional cylindrical equation. To avoid the effect of these oscillations, $l_0$, $L_x$ and correspondingly the number of grid points in $x$ need to increase for better agreement.

\section{Conclusion}
In this paper we consider DSW behavior and exact reductions across a parabolic front of $2+1$ dimensional KP and 2DBO equations to the $1+1$ dimensional cylindrical KdV (cKdV) and the $1+1$ dimensional cylindrical BO (cBO) equations. We derive their associated modulation equations and write these equations in suitable Riemann coordinates. We solve the resulting Whitham systems numerically and compare these results with direct numerical simulations of the $1+1$ equations. Apart from an unimportant phase and a small discontinuity in front of the DSW, the results are in excellent agreement. The discontinuity can be accounted for by higher order terms; but this detail is outside the scope of the present paper.  We also compare the Whitham theory of cKdV with KdV and cBO with BO; we find that while the amplitudes of the DSW structures of KdV and BO remain $O(1)$, the DSWs of cKdV and cBO decay slowly in time. We compare the DSW behavior across the parabolic front of the $2+1$ systems to their $1+1$ counterparts; after accounting for an  small mean term; excellent agreement is obtained. We conclude that the cKdV/cBO equations are able to accurately describe DSW behavior along a `flattening' parabolic front to the KP/2DBO equations.

\section{Acknowledgements}
This research was partially supported by the U.S. Air Force Office of Scientific Research, under grant FA9550-12-1-0207 and by the NSF under grant DMS-1310200 and by the Scientific and Technological Research Council of Turkey (TUBITAK) under grant 1059B-19140044.

\newpage

\bibliographystyle{elsarticle-num}
\biboptions{compress}

\end{document}